\documentclass[aip,graphicx,amsmath,amssymb,reprint]{revtex4-1}
\usepackage{physics}
\usepackage{times}
\usepackage{graphicx}
\usepackage{amsfonts}
\usepackage{amsmath, amsthm, amssymb, mathrsfs}
\usepackage{empheq}
\usepackage{dsfont}
\usepackage{xcolor}
\usepackage{cancel}
\usepackage{standalone}
\usepackage{stmaryrd}
\usepackage{microtype}
\usepackage{tikz}
\usetikzlibrary{decorations.markings}
\usetikzlibrary{shapes,positioning,arrows}
\usepackage{bm}
\usepackage{siunitx}
\usepackage{slashed}
\usepackage{siunitx}\usepackage[colorlinks=true, allcolors=black, citecolor=blue, linkcolor=blue, urlcolor=blue]{hyperref}
\usepackage[capitalize]{cleveref}
\usepackage{epsfig}

\usepackage{soul}

\allowdisplaybreaks

\usepackage{graphicx}
\usepackage[caption=false]{subfig}
\graphicspath{{figures/}}

\newcommand{\del}{\partial}

\newcommand{\g}{\gamma}
\newcommand{\gt}{\tilde{\gamma}}

\newcommand{\vv}[1]{{\mathbf{#1}}}
\renewcommand{\vb}{\bm}

\newcommand{\gR}{\hat{g}}

\newcommand{\s}{\sigma}
\newcommand{\vs}{\bm{\s}}

\newcommand{\T}{\hat{\mathcal{T}}}

\newcommand{\B}{\hat{\mathcal{B}}}

\newcommand{\basis}{\vv{e}}

\DeclareMathAlphabet{\mathpzc}{OT1}{pzc}{m}{it}

\newcommand{\eg}{\emph{e.g.}~}

\newcommand{\be}{\begin{equation}}
\newcommand{\ee}{\end{equation}}
\newcommand{\bse}{\begin{subequations}}
\newcommand{\ese}{\end{subequations}}
\newcommand{\bal}{\begin{align}}
\newcommand{\eal}{\end{align}}

\newcommand{\curvba}[1]{\hat{\bm{\mathcal{#1}}}}

\usepackage[integrals]{wasysym}

\begin{document}

\title{Electrical control of superconducting spin valves using ferromagnetic helices}

\author{Tancredi Salamone}
\affiliation{Center for Quantum Spintronics, Department of Physics, NTNU, Norwegian University of Science and Technology, NO-7491 Trondheim, Norway}
\author{Henning G. Hugdal}
\affiliation{Center for Quantum Spintronics, Department of Physics, NTNU, Norwegian University of Science and Technology, NO-7491 Trondheim, Norway}
\author{Morten Amundsen}
\affiliation{Center for Quantum Spintronics, Department of Physics, NTNU, Norwegian University of Science and Technology, NO-7491 Trondheim, Norway}
\author{Sol H. Jacobsen}
\email[Corresponding author: ]{sol.jacobsen@ntnu.no}
\affiliation{Center for Quantum Spintronics, Department of Physics, NTNU, Norwegian University of Science and Technology, NO-7491 Trondheim, Norway}

\begin{abstract}
     The geometrical properties of a helical ferromagnet are shown theoretically to control the critical temperature of a proximity-coupled superconductor. Using the Usadel equation for diffusive spin transport, we provide self-consistent analysis of how curvature and torsion modulate the proximity effect. When the helix is attached to a piezoelectric actuator, the pitch of the helix -- and hence the superconducting transition -- can be controlled electrically. 
\end{abstract}


\maketitle
Control of superconductivity is central to the ambitious aims of low-dissipation computing alternatives in superconducting spintronics \cite{Eschrig2011,Linder2015}. In superconducting spin valve devices, a tunable parameter controls the superconducting critical temperature ($T_c$), toggling high and low resistance states. It is well established that $T_c$ of a conventional singlet superconductor can be controlled by the relative magnetization orientation of proximity-coupled magnetic multilayers, for example in superconductor-ferromagnet-ferromagnet (SFF') or FSF' configurations \cite{Oh1997,Tagirov1999,Buzdin1999,Gu2002,Potenza2005,Moraru2006,Leksin2012,Wang2014,Zdravkov2013,Banerjee2014}. The different alignments of the magnetic multilayers provide competing spin orientation axes. When magnets are proximity-coupled to a superconductor, their relative magnetization angle provides a mechanism to drain the superconductor of singlet Cooper pairs, by converting them into triplet pairs in the magnets \cite{Bergeret2001,Khaire2010,Robinson2010}. As the number of singlet pairs decreases in the superconductor, so does its $T_c$.

Having multiple magnetic interfaces introduces several sources of spin scattering, which reduces overall device efficiency, and precise control of magnetic misalignment in multilayers is challenging \cite{Leksin2012,Wang2014,Zdravkov2013,Banerjee2014,Banerjee2018}. Using the same principle of combining misaligned $\text{SU}(2)$ fields to convert between singlets and triplets \cite{Bergeret2013,Bergeret2014,amundsen2022}, $T_c$ can instead be modulated by the relative strengths of Rashba and Dresselhaus spin-orbit coupling (SOC) in a single ferromagnetic layer for an SF valve \cite{Jacobsen2015b}. This has recently been shown experimentally, for a Nb superconductor proximity coupled to an asymmetric Pt/Co/Pt trilayer \cite{Banerjee2018}, where the angle of applied magnetic field controls the triplet conversion by modulating the relationship between the exchange field and overall SOC-vector, and hence gives some magnetic control over $T_c$. It has also been proposed that the relative proportions of Rashba and Dresselhaus SOC can be tuned electrically via voltage gating\cite{Ouassou2016}. These proposals mitigate the problem of spin scattering occurring at multiple interfaces, and remove the difficult problem of precision in controlling the misalignment angle between multiple magnetic layers. However, the Rashba-Dresselhaus SOC-profile places rather strict restrictions on material choice. In contrast, a recent alternative to Rashba-Dresselhaus SOC-control is to design and control the SOC-profile of magnetic materials directly through its geometric curvature profile\cite{Salamone2021,Salamone2022,Skarpeid2024,Salamone2024}.
\begin{figure}
    \centering
    \includegraphics[width=\columnwidth]{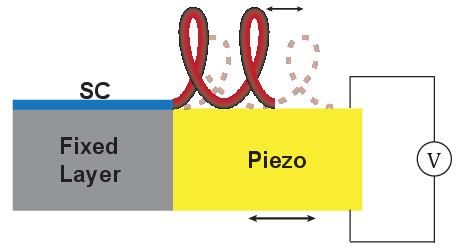}
    \caption{A superconducting nanowire is proximity-coupled to a ferromagnetic helix, where the pitch of the helix is controlled electrically with a piezoelectric actuator.}
    \label{fig:SF_helix}
\end{figure}

There is booming interest in geometric control in spintronics and magnetism \cite{Gentile2022,Streubel2016,Streubel2021,Gentile2013,Sheka2022,Makarov2022}, and, since real-space modulation rotates the spin quantization axis, curvature has emerged as an alternative route to designing and controlling the effective SOC throughout a magnet \cite{Salamone2021,Salamone2022,Skarpeid2024,Salamone2024}. This can for example be employed to control the direction of Josephson current \cite{Salamone2021}, probe the quality of magnetic ordering at buried interfaces between superconductors and antiferromagnets \cite{Salamone2024}, and using the inherent structural chirality to probe mixed-chirality junctions \cite{Skarpeid2024}. We have previously shown how in-plane curvature of an SF bilayer can modulate the critical temperature of the superconductor \cite{Salamone2022}, and explored how it is affected by the presence of both curvature-induced and intrinsic SOC. Here, the additional effect of torsion is investigated, and we highlight that for tangential exchange fields the roles of curvature and torsion differ. Due to a large response in the critical temperature with small changes in torsion, helices with tangential exchange fields are well suited to piezoelectric actuator control.

The experimental field of curvilinear magnetism employs a range of fabrication techniques to create intricate nanoscale structures and arrays, including electron-beam lithography \cite{Volkov2019}, two-photon lithography \cite{Sahoo2018}, focused-electron beam induced deposition \cite{Dobrovolskiy2021,Sanz-Hernandez2020,Skoric2020}, and glancing angle deposition \cite{Gibbs2014}, which has produced ferromagnetic helices (e.g. nickel) with a radius down to approximately $10$~nm. Inspired by a proposal for piezoelectric switching between helimagnetic states \cite{Volkov_2019_JPhysD}, we will show that a ferromagnetic helix can be used to manufacture a superconducting spin valve that can be controlled electrically, by employing a piezoelectric actuator, which controls piezoelectric strain response via an applied voltage (see \cref{fig:SF_helix}). A sample is fixed to the piezoelectric via, for example, gold thermo-compression bonding – used in controlling nanomembrane LEDs \cite{Trotta2012} – or an insulating hardened glue such as cyanoacrylate or polymethyl methacrylate – used in controlling quantum dots embedded in a strainable sample \cite{Plumhof2011,Ding2010,Zander2009}. We propose a similar method can affix and control the strain in the helix in \cref{fig:SF_helix}. Note the magnetic helix has no intrinsic SOC, and no external field is applied. We have instead fully electrical control of the helix geometry, which we will show to give significant control over the superconducting transition.

\begin{figure}
    \centering
    \includegraphics[width=.7\columnwidth]{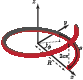}
    \caption{Characteristic parameters of a helix, with radius $R$, arclength $s$, pitch $2c\pi$, and azimuthal angle $\phi$.}
    \label{fig:helix_param}
\end{figure}
Helix growth via glancing angle deposition, for example, renders the wire inherently diffusive, and hence we employ the quasiclassical framework for diffusive spin transport, i.e. the Usadel equation \cite{Usadel1970} and associated boundary conditions \cite{KuprianovLukichev1988}.  These are transformed into curvilinear coordinates via the Frenet-Serret framework\cite{ortix2015quantum}, where the three-dimensional space around the helix in \cref{fig:helix_param} is parameterized as $\vb{R}(s, n, b) = \vb{r}(s) + \curvba{N}(s) n + \curvba{B}(s) b$, with $\vb{r}(s)$ parametrizing the curve along the arclength $s$, and $n$ and $b$ representing the normal and binormal coordinates, respectively. For a nanowire helix, $n, b \rightarrow 0$. Orthogonal unit vectors in the tangential, normal and binormal curvilinear directions are respectively $\curvba{T}(s) = \partial_s \vb{r}(s)$, $\curvba{N}(s) = \partial_s \curvba{T}(s)/\left|\del_s\T(s)\right|$ and $\curvba{B}(s) = \curvba{T}(s) \times \curvba{N}(s)$ , indicated in \cref{fig:helix_param}. The curvature $\kappa(s)=|\del_s\T(s)|$ and torsion $\tau(s)=|\del_s\B(s)|$ are then connected through the Frenet-Serret formulas \cite{ortix2015quantum}
\begin{equation}
    \begin{pmatrix} \partial_s \curvba{T}(s) \\ \partial_s \curvba{N}(s) \\ \partial_s \curvba{B}(s)  \end{pmatrix} = \begin{pmatrix} 0 & \kappa(s) & 0 \\ -\kappa(s) & 0 & \tau(s) \\ 0 & -\tau(s) & 0 \end{pmatrix} \begin{pmatrix} \curvba{T}(s) \\ \curvba{N}(s) \\ \curvba{B}(s) \end{pmatrix}.
\end{equation}

A helical nanowire with $n_t$ turns, radius $R$ and $2c\pi$ pitch, i.e. height of a complete helix turn, can be parametrized by\cite{ortix2015quantum} the azimuthal angle $\phi=\left[0,2n_t\pi\right]$, with
\bse\begin{align}
	x &= R\cos\phi,\\
	y &= R\sin\phi,\\
	z &= c\phi,
\end{align}\ese
\noindent and $c$ determines the out-of-plane tilt of the nanowire, shown in \cref{fig:helix_param}. Curvature and torsion are respectively given by 
\bse\begin{align}
\kappa&=R/(R^2+c^2) , \label{eq:kappa_exp}\\
\tau&=c/(R^2+c^2). \label{eq:tau_exp}
\end{align}\ese

\noindent The arclength coordinate is given by $s=\phi\sqrt{R^2+c^2}$. This parametrization leads to the three unit vectors:
\begin{subequations}
\begin{align}
    \curvba{T}(s) &= -\cos\alpha\sin\phi\hat{\basis}_x + \cos\alpha\cos\phi\hat{\basis}_y + \sin\alpha\hat{\basis}_z, \\
    \curvba{N}(s) &= -\cos\phi\hat{\basis}_x - \sin\phi\hat{\basis}_y, \\
    \curvba{B}(s) &= \sin\alpha\sin\phi\hat{\basis}_x - \sin\alpha\cos\phi\hat{\basis}_y + \cos\alpha\hat{\basis}_z,
\end{align}
\end{subequations}
where $\alpha = \arctan{(\tau/\kappa)}$. The curvilinear Pauli matrices are then 
\begin{equation} \label{eq:Curvilinear Pauli matrices}
    \sigma_T = \bm{\sigma} \cdot \curvba{T}(s), \quad \sigma_N = \bm{\sigma} \cdot \curvba{N}(s), \quad \sigma_B = \bm{\sigma} \cdot \curvba{B}(s),
\end{equation}
with Pauli vector $\bm{\sigma}=(\sigma_T,\sigma_N,\sigma_B)$. In this way, the Pauli matrices can encode the exchange-field dependence on the system geometry. 

To solve the curvilinear Usadel equation numerically, we use the Riccati parametrization of the quasiclassical Green's function \cite{Jacobsen2015b,Schopohl1995}:
\begin{equation} \label{eq:Riccati parametrization}
    \gR = \begin{pmatrix} N & 0 \\ 0 & -\Tilde{N} \end{pmatrix} \begin{pmatrix} 1 + \gamma \tilde{\gamma} & 2 \gamma \\ 2 \Tilde{\gamma} & 1 + \Tilde{\gamma} \gamma \end{pmatrix},
\end{equation}
where $N$ is a $2 \times 2$ normalization matrix defined as $N = (1 - \gamma \Tilde{\gamma})^{-1}$. The tilde operation is defined as $\Tilde{\gamma}(s, \varepsilon) = \gamma^*(s, -\varepsilon)$. 
The Riccati parameterized Usadel equation thus becomes:
\begin{equation} \label{eq:Parametrized AF Usadel}
     iD[\partial_s^2 \gamma\!+\!2(\partial_s \gamma)\Tilde{N}\Tilde{\gamma}(\partial_s \gamma)]=2\varepsilon\gamma\!+\!(\bm{h}\cdot\vs\g\!-\!\g\bm{h}\cdot\vs^*),
\end{equation}
where $D$ is the diffusion constant and $\bm{h}$ is the exchange field.
Similarly, the curvilinear Kupriyanov-Lukichev\cite{KuprianovLukichev1988} boundary conditions become\cite{Salamone2022}:
\begin{subequations}\label{eq:BCs}
\begin{align}
    L_S\zeta_S\bar{\partial}_s \gamma_S = (1-\gamma_S\gt_F)N_F(\gamma_F - \gamma_S),\\
    L_F\zeta_F\bar{\partial}_s \gamma_F = (1-\gamma_F\gt_S)N_S(\gamma_F - \gamma_S),
\end{align}
\end{subequations}
with length $L_{S(F)}$, and interface parameter $\zeta_{S(F)}$ of the superconducting (ferromagnetic) region representing the ratio of barrier to bulk resistance. The term $\bar{\del}_s$ denotes the derivative evaluated at the interface. 

We take the bulk superconductor to have spin-singlet gap $\Delta_0$, and corresponding bulk critical temperature $T_{c0}$. Proximity-coupling to a ferromagnet enables the conversion of bulk spin-singlet pairs into odd-frequency spin-triplets, that leak into the ferromagnet, spreading superconductivity through the system and weakening the superconductor. By controlling the interconversion between triplet species, these may penetrate further into the magnet and drain even more singlets from the superconductor. The critical temperature is governed by the density of the singlet order parameter, and we can calculate this self-consistently for any particular SF combination.

Numerically, we calculate the critical temperature by finding a self-consistent solution to the Usadel \cref{eq:Parametrized AF Usadel}, using boundary conditions \cref{eq:BCs} and the gap equation\cite{Jacobsen2015b}
\begin{eqnarray}\label{Eq:gap}
    \Delta(s,T) = N_0 \lambda \int_{0}^{\omega_c}d\epsilon \textrm{Re}\{f_0(\epsilon,s)\}\textrm{tanh}\left(\frac{\pi}{2e^\gamma}\frac{\epsilon/\Delta_0}{T/T_{c0}}\right)\!\!.
\end{eqnarray}
Here $\lambda$ is the coupling constant between electrons, $N_0$ is the density of states at the Fermi level, $\omega_c=\Delta_0\textrm{cosh}(1/N_0\lambda)$ is the cut-off energy, $\gamma\simeq 0.577$ is the Euler-Mascheroni constant, and $T$ is the temperature. We use the binary search algorithm presented in \cite{Ouassou2016}, which is openly accessible on GitHub as part of the GENEUS set of numerical programs \cite{Geneus}, modified to include geometric curvature as presented above.

\begin{figure}
    \centering
    \includegraphics[width=\columnwidth]{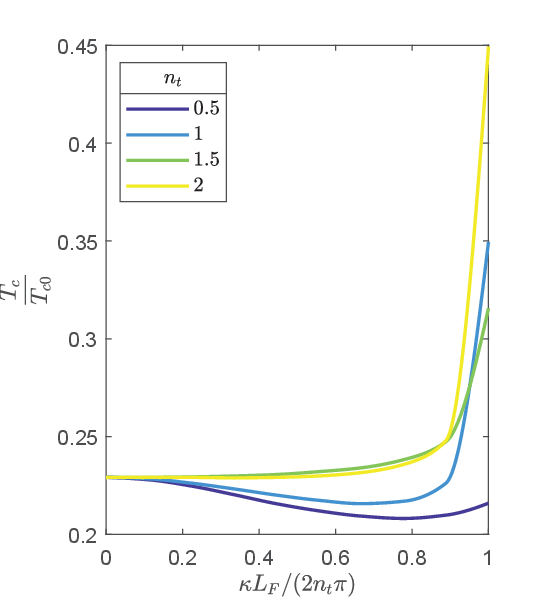}
    \caption{Critical temperature of the superconducting-ferromagnetic helix hybrid nanowire for $n_t=0.5$, 1, 1.5 and 2 number of turns, with $L_S=0.55\xi_S$, $L_F=0.5\xi_S$, $\bm{h}=10\Delta_0\T$. }
    \label{fig:Tc_vs_k_n1_nins.5}
\end{figure}

Using the binary search algorithm, we calculate $\Delta(s,T)$ for $M$ different values of $T$, which gives a numerical estimate for the critical temperature with a precision of $T_{c0}/2^{M+1}$. The numerical values presented here are obtained for $M=15$. We show this critical temperature in \cref{fig:Tc_vs_k_n1_nins.5} for different helices of fixed length in the range $n_t=0.5-2$. In each case, the total length of the helical wire is constant, $L_F=0.5\xi_S$ for this example, which means that the curvature and torsion have a fixed dependency, \eg if we fix $\kappa$, $\tau$ is given by $\tau=\sqrt{(2\pi n_t/L_{F})^2-\kappa^2}$, with $\kappa\leq2\pi n_t/L_{F}$. The plots in \cref{fig:Tc_vs_k_n1_nins.5} are the result of an interpolation of the values of $T_c$ obtained from 10 different $\kappa$, $\tau$ pairs. A quick calculation suggests that to have a ferromagnetic helix of radius $R\approx 10$ nm, which is the approximate limit of current experimental capabilities, the superconducting coherence length should be of the order of tens-to-hundreds of nanometers. Since the diffusive coherence length can be estimated $\xi_S=\sqrt{l \xi_0}$, where $l$ is the mean free path and $\xi_0$ the ballistic superconducting coherence length, we can see that such values are easily achievable for many superconducting materials (e.g. aluminium has $\xi_0\approx1600$ nm). We note that the mean free path is comparable to the smallest system dimension\cite{Zgirski2007,Arutyunov2008}, in this case the wire diameter, which would be of the order of a few nanometers for the smallest helices. However, the helix radius, wire length and thickness can all be adjusted based on experimental requirements, within the applicability of dominant transport in one direction.

To provide some analytic insight into the role of curvature and torsion, we consider the limit of weak proximity coupling, where the components of the $\gamma$-matrix are expected to be small, i.e. $\abs{\gamma_{ij}} \ll 1$, such that we may neglect terms of the order $\mathcal{O}(\gamma^2)$, and take $N \approx 1$. The anomalous Green's function given in the upper right block of equation \eqref{eq:Riccati parametrization} becomes $f = 2\gamma$, and is describable by a scalar singlet component $f_0$, and a vector of triplet components, encapsulated in the so-called $\bm{d}$-vector,
\begin{equation}\label{Eq:dvec}
    f = (f_0 + \bm{d} \cdot \bm{\sigma})i\tau_2,
\end{equation}
where $\bm{d} = (d_T, d_N, d_B)$, and $\tau_2=\mathrm{antidiag}(-i,i)$. In this limit, we can re-express the Usadel equations in $\bm{d}$-vector form, and capture the primary principles controlling the diffusive transport of superconductivity. The weak proximity equations take the form
\bse
\label{eq2:wp_F}
\begin{align}
    \frac{iD}{2}&(\del^2_sd_T\!-\!2\kappa\del_sd_N)\!=\!f_0h_T\!+\!\left(\!\varepsilon\!+\frac{i}{2}D\kappa^2\!\right)\!d_T\!-\!\frac{i}{2}D\kappa\tau d_B, \label{eq2:wpdT_F}\\
    \frac{iD}{2}&(\del^2_sd_N\!+\!2\kappa\del_sd_T-2\tau\del_sd_B)\nonumber\\*
    &\hspace{1.5cm}=\!f_0h_N +\left[\varepsilon\!+\frac{i}{2}D(\kappa^2\!+\!\tau^2)\right]\!d_N, \label{eq2:wpdN_F}\\
    \frac{iD}{2}&(\del^2_sd_B\!+\!2\tau\del_sd_N)\!=\!f_0h_B\!+\!\left(\!\varepsilon\!+\!\frac{i}{2}\!D\tau^2\!\!\right)\!d_B\!-\!\frac{i}{2}D\kappa\tau d_T, \label{eq2:wpdB_F}\\
    \frac{iD}{2}&\del^2_sf_0\!=\!\varepsilon f_0+\bm{h}\cdot\bm{d}. \label{eq2:wpf0_F}
\end{align}
\ese

We see first, in \cref{eq2:wpf0_F}, that the exchange field converts singlets into triplet pairings. The effect of $\kappa$ and $\tau$ in \crefrange{eq2:wpdT_F}{eq2:wpdB_F} is twofold, dictating how the triplets undergo spin-precession and spin-relaxation. The former can be identified with first derivative terms, proportional to $\kappa$ and $\tau$, and describes the rotation of the spin of superconducting triplet correlations while moving along the ferromagnet. The latter appears as an additional imaginary component of the triplet energy proportional to $\kappa^2$ and $\tau^2$, and represents a loss of spin information due to frequent impurity scattering. Through their role in spin precession, the curvature and torsion can independently rotate the triplets to gain finite spin polarization, which can cause superconductivity to penetrate further into the magnetic material and contribute to draining more singlets from the superconductor. At the same time, large values of curvature and torsion will have an increased spin relaxation effect, and reduce the overall magnitude of the proximity coupling.

Considering specifically the case of a tangential exchange field, which is a simplified but typically assumed ground state of many ferromagnetic nanowires \cite{Tkachenko2012,Tkachenko2013,Slastikov2011}, we can note that spin-relaxation would be more effective for high values of $\kappa$, rather than high values of $\tau$. This can be explained as follows. The tangential exchange field generates tangential triplets ($d_T$), which are then converted into normal ($d_N$) and binormal ($d_B$) triplets through the effect of curvature and torsion. Tangential triplets are mainly affected by $\kappa$ for both spin-precession and spin-relaxation. Therefore, when $\kappa$ is high, spin-relaxation ($\sim\kappa^2$) dominates over spin-precession ($\sim\kappa$). This rapid dominance at high $\kappa$ gives us a the large change in $T_c$, making this parameter range perfectly suited for spin-valve control through a small applied strain. On the other hand, when $\kappa$ is small, the conversion of tangential triplets to normal and binormal triplets is reduced. The spin-relaxation due to $\tau$ would instead be significant, but it mainly affects normal and binormal triplets, which are converted in small amounts with a tangential exchange field. 
For ground state exchange fields with a different orientation, the roles of $\kappa$ and $\tau$ would differ from this description, still following \cref{eq2:wp_F}.

Examining \cref{fig:Tc_vs_k_n1_nins.5}, we see that for $n_t=0.5$, there is considerable variation in $T_c$ for all $\kappa$. Since the length is kept fixed, larger $n_t$ at $\tau=0$ corresponds to a smaller radius of curvature, which is an increase in the effective SOC. The $n_t=0.5$ curve thus has the smallest effective SOC, and spin precession remains significant for all $\kappa$, dominating when $\kappa \tau$ and $\tau/L_F$ are larger than $\kappa/L_F$ and $\kappa^2$. This explains the initial decrease in $T_c$, since conversion between $d_T$ and $d_B$, with subsequent relaxation and rotation to $d_N$, drains the $d_T$ channel and thus also the singlets. As the number of turns increases, there is an effective increase in curvature-induced SOC, and spin diffusion becomes more dominant. For high curvatures, $\kappa L_F/2n_t\pi\sim 0.8-1$, we still see an influence of spin precession in the nonmonotonicity in the $\tau=0$ limit, evidence of the SOC-based oscillations in $T_c$ as expected\cite{Salamone2022}. However, there is a sharp increase in $T_c$ as the triplet-conversion pathway deteriorates, restoring $T_c$ towards its bulk value. We can utilize this sharp response to small changes in the geometric curvature/torsion relationship to create an effective superconducting spin valve. 

A number of different mechanisms can be employed to tune the helical geometry in-situ, for example via mechanical strain, or strain induced by thermoelectricity or light \cite{Kundys2015,Matzen2019,Guillemeney2022}. To suggest a specific implementation, we propose a method for electrical control of the geometry through strain induced in a piezoelectric material, as illustrated in \cref{fig:SF_helix}. Achievable strain in piezoelectric ceramic materials is typically around $0.2\%$, and over $1\%$ in single crystals \cite{Park1997}. We can estimate the strain-induced change in $T_c$ by relating the change in pitch to the change in torsion from \cref{eq:tau_exp}, and its corresponding change in curvature $\kappa=\sqrt{(2\pi n_t/L_F)^2-\tau^2}$. The variation in $T_c$ is then calculated through an interpolation of the numerical results as a function of $\kappa$, reported in \cref{fig:Tc_vs_k_n1_nins.5}. Taking a conservative value of strain of $0.2\%$, the resultant change in pitch of the helix would correspond to a change of about $0.05\%$ in $\kappa L_F/2n_t \pi$. For a single turn of the helix near the point of steepest ascent, this would result in a change in $T_c$ of more than $1\%$, which is easily measurable with currently available techniques (see e.g. \cite{Banerjee2014}). For strains of order $1\%$, the corresponding change in $T_c$ would be of order $3\%$.


 In our analysis we have assumed a magnetic ground state where the exchange field remains tangential regardless of geometric configuration. This is reasonable for most choices of ferromagnetic wires with in-plane curvatures below a critical threshold\cite{Sheka2015,Salamone2022}. When introducing torsion, the exchange field remains tangential if the $\kappa < \kappa_b$, where $\kappa_b$ is a $\tau$-dependent critical curvature\cite{Sheka2015a}. Between these parameter ranges, the exchange field acquires a tilt into the binormal direction, and investigation of this regime would be interesting for further study. Keeping a tangential exchange field for simplicity here, where we have a helix with a small pitch, we see that the primary parameter range of interest is the region $\kappa L_F/2n_t\pi \sim 0.8-1$. We thus keep $n_t\leq 2$, to ensure the tangential exchange field ground state remains valid for most materials with $L_F = 0.5\xi_S$ \cite{Salamone2022}. However, notice also that the number of turns has an even greater effect on the change in $T_c$ than changing $\kappa$ and $\tau$ alone. This suggests that a control mechanism that increases the number of helical turns instead of the helical pitch would be even more effective in controlling the superconducting transition.
 
 The one-dimensional system we present here could be parallelized on a thin film superconductor, for example via adsorption of helical magnetic molecules on a superconducting substrate, or potentially also for island arrays of non-magnetic chiral molecules that collectively behave magnetically, for which a drop in $T_c$ has been reported\cite{Ozeri2023}. The more tightly packed the array of magnetic helices is, the stronger is their collective ability to drain singlets from the superconductor and control the thin-film $T_c$. 
 
We argue that the future of superconducting spintronics stands to benefit significantly from recent advances in curvilinear magnetism, bringing with it a range of new design and control mechanisms. Here we have proposed a specific design for a superconducting spin valve, controlled electrically via the strain response through a piezoelectric substrate. Several alternative mechanisms for controlling the strain in-situ abound, and it would be very interesting to explore alternative designs that employ instead a photostrictive substrate, or materials with a thermoelectric or large mechanical strain response.

\begin{acknowledgments}
We thank D. Makarov for helpful comments. Computations have been performed on the SAGA supercomputer provided by UNINETT Sigma2 - the National Infrastructure for High Performance Computing and Data Storage in Norway. We acknowledge funding via the “Outstanding Academic Fellows” programme at NTNU, and the Research Council of Norway Grant Nos. 302315 and 262633.
\end{acknowledgments}

\bibliography{references}

\begin{thebibliography}{59}%
\makeatletter
\providecommand \@ifxundefined [1]{%
 \@ifx{#1\undefined}
}%
\providecommand \@ifnum [1]{%
 \ifnum #1\expandafter \@firstoftwo
 \else \expandafter \@secondoftwo
 \fi
}%
\providecommand \@ifx [1]{%
 \ifx #1\expandafter \@firstoftwo
 \else \expandafter \@secondoftwo
 \fi
}%
\providecommand \natexlab [1]{#1}%
\providecommand \enquote  [1]{``#1''}%
\providecommand \bibnamefont  [1]{#1}%
\providecommand \bibfnamefont [1]{#1}%
\providecommand \citenamefont [1]{#1}%
\providecommand \href@noop [0]{\@secondoftwo}%
\providecommand \href [0]{\begingroup \@sanitize@url \@href}%
\providecommand \@href[1]{\@@startlink{#1}\@@href}%
\providecommand \@@href[1]{\endgroup#1\@@endlink}%
\providecommand \@sanitize@url [0]{\catcode `\\12\catcode `\$12\catcode `\&12\catcode `\#12\catcode `\^12\catcode `\_12\catcode `\%12\relax}%
\providecommand \@@startlink[1]{}%
\providecommand \@@endlink[0]{}%
\providecommand \url  [0]{\begingroup\@sanitize@url \@url }%
\providecommand \@url [1]{\endgroup\@href {#1}{\urlprefix }}%
\providecommand \urlprefix  [0]{URL }%
\providecommand \Eprint [0]{\href }%
\providecommand \doibase [0]{http://dx.doi.org/}%
\providecommand \selectlanguage [0]{\@gobble}%
\providecommand \bibinfo  [0]{\@secondoftwo}%
\providecommand \bibfield  [0]{\@secondoftwo}%
\providecommand \translation [1]{[#1]}%
\providecommand \BibitemOpen [0]{}%
\providecommand \bibitemStop [0]{}%
\providecommand \bibitemNoStop [0]{.\EOS\space}%
\providecommand \EOS [0]{\spacefactor3000\relax}%
\providecommand \BibitemShut  [1]{\csname bibitem#1\endcsname}%
\let\auto@bib@innerbib\@empty
\bibitem [{\citenamefont {Eschrig}(2011)}]{Eschrig2011}%
  \BibitemOpen
  \bibfield  {author} {\bibinfo {author} {\bibfnamefont {M.}~\bibnamefont {Eschrig}},\ }\bibfield  {title} {\enquote {\bibinfo {title} {Spin-polarized supercurrents for spintronics},}\ }\href {\doibase 10.1063/1.3541944} {\bibfield  {journal} {\bibinfo  {journal} {Physics Today}\ }\textbf {\bibinfo {volume} {64}},\ \bibinfo {pages} {43--49} (\bibinfo {year} {2011})}\BibitemShut {NoStop}%
\bibitem [{\citenamefont {Linder}\ and\ \citenamefont {Robinson}(2015)}]{Linder2015}%
  \BibitemOpen
  \bibfield  {author} {\bibinfo {author} {\bibfnamefont {J.}~\bibnamefont {Linder}}\ and\ \bibinfo {author} {\bibfnamefont {J.~W.~A.}\ \bibnamefont {Robinson}},\ }\bibfield  {title} {\enquote {\bibinfo {title} {Superconducting spintronics},}\ }\href {\doibase 10.1038/nphys3242} {\bibfield  {journal} {\bibinfo  {journal} {Nature Physics}\ }\textbf {\bibinfo {volume} {11}},\ \bibinfo {pages} {307--315} (\bibinfo {year} {2015})}\BibitemShut {NoStop}%
\bibitem [{\citenamefont {Oh}, \citenamefont {Youm},\ and\ \citenamefont {Beasley}(1997)}]{Oh1997}%
  \BibitemOpen
  \bibfield  {author} {\bibinfo {author} {\bibfnamefont {S.}~\bibnamefont {Oh}}, \bibinfo {author} {\bibfnamefont {D.}~\bibnamefont {Youm}}, \ and\ \bibinfo {author} {\bibfnamefont {M.~R.}\ \bibnamefont {Beasley}},\ }\bibfield  {title} {\enquote {\bibinfo {title} {{A superconductive magnetoresistive memory element using controlled exchange interaction}},}\ }\href {\doibase 10.1063/1.120032} {\bibfield  {journal} {\bibinfo  {journal} {Applied Physics Letters}\ }\textbf {\bibinfo {volume} {71}},\ \bibinfo {pages} {2376--2378} (\bibinfo {year} {1997})}\BibitemShut {NoStop}%
\bibitem [{\citenamefont {Tagirov}(1999)}]{Tagirov1999}%
  \BibitemOpen
  \bibfield  {author} {\bibinfo {author} {\bibfnamefont {L.~R.}\ \bibnamefont {Tagirov}},\ }\bibfield  {title} {\enquote {\bibinfo {title} {Low-field superconducting spin switch based on a superconductor $/$ferromagnet multilayer},}\ }\href {\doibase 10.1103/PhysRevLett.83.2058} {\bibfield  {journal} {\bibinfo  {journal} {Phys. Rev. Lett.}\ }\textbf {\bibinfo {volume} {83}},\ \bibinfo {pages} {2058--2061} (\bibinfo {year} {1999})}\BibitemShut {NoStop}%
\bibitem [{\citenamefont {Buzdin}, \citenamefont {Vedyayev},\ and\ \citenamefont {Ryzhanova}(1999)}]{Buzdin1999}%
  \BibitemOpen
  \bibfield  {author} {\bibinfo {author} {\bibfnamefont {A.~I.}\ \bibnamefont {Buzdin}}, \bibinfo {author} {\bibfnamefont {A.~V.}\ \bibnamefont {Vedyayev}}, \ and\ \bibinfo {author} {\bibfnamefont {N.~V.}\ \bibnamefont {Ryzhanova}},\ }\bibfield  {title} {\enquote {\bibinfo {title} {Spin-orientation–dependent superconductivity in {F/S/F} structures},}\ }\href {\doibase 10.1209/epl/i1999-00539-0} {\bibfield  {journal} {\bibinfo  {journal} {Europhysics Letters}\ }\textbf {\bibinfo {volume} {48}},\ \bibinfo {pages} {686} (\bibinfo {year} {1999})}\BibitemShut {NoStop}%
\bibitem [{\citenamefont {Gu}\ \emph {et~al.}(2002)\citenamefont {Gu}, \citenamefont {You}, \citenamefont {Jiang}, \citenamefont {Pearson}, \citenamefont {Bazaliy},\ and\ \citenamefont {Bader}}]{Gu2002}%
  \BibitemOpen
  \bibfield  {author} {\bibinfo {author} {\bibfnamefont {J.~Y.}\ \bibnamefont {Gu}}, \bibinfo {author} {\bibfnamefont {C.-Y.}\ \bibnamefont {You}}, \bibinfo {author} {\bibfnamefont {J.~S.}\ \bibnamefont {Jiang}}, \bibinfo {author} {\bibfnamefont {J.}~\bibnamefont {Pearson}}, \bibinfo {author} {\bibfnamefont {Y.~B.}\ \bibnamefont {Bazaliy}}, \ and\ \bibinfo {author} {\bibfnamefont {S.~D.}\ \bibnamefont {Bader}},\ }\bibfield  {title} {\enquote {\bibinfo {title} {Magnetization-orientation dependence of the superconducting transition temperature in the ferromagnet-superconductor-ferromagnet system: $\mathrm{C}\mathrm{u}\mathrm{N}\mathrm{i}/\mathrm{N}\mathrm{b}/\mathrm{C}\mathrm{u}\mathrm{N}\mathrm{i}$},}\ }\href {\doibase 10.1103/PhysRevLett.89.267001} {\bibfield  {journal} {\bibinfo  {journal} {Phys. Rev. Lett.}\ }\textbf {\bibinfo {volume} {89}},\ \bibinfo {pages} {267001} (\bibinfo {year} {2002})}\BibitemShut {NoStop}%
\bibitem [{\citenamefont {Potenza}\ and\ \citenamefont {Marrows}(2005)}]{Potenza2005}%
  \BibitemOpen
  \bibfield  {author} {\bibinfo {author} {\bibfnamefont {A.}~\bibnamefont {Potenza}}\ and\ \bibinfo {author} {\bibfnamefont {C.~H.}\ \bibnamefont {Marrows}},\ }\bibfield  {title} {\enquote {\bibinfo {title} {Superconductor-ferromagnet $\mathrm{CuNi}/\mathrm{Nb}/\mathrm{CuNi}$ trilayers as superconducting spin-valve core structures},}\ }\href {\doibase 10.1103/PhysRevB.71.180503} {\bibfield  {journal} {\bibinfo  {journal} {Phys. Rev. B}\ }\textbf {\bibinfo {volume} {71}},\ \bibinfo {pages} {180503} (\bibinfo {year} {2005})}\BibitemShut {NoStop}%
\bibitem [{\citenamefont {Moraru}, \citenamefont {Pratt},\ and\ \citenamefont {Birge}(2006)}]{Moraru2006}%
  \BibitemOpen
  \bibfield  {author} {\bibinfo {author} {\bibfnamefont {I.~C.}\ \bibnamefont {Moraru}}, \bibinfo {author} {\bibfnamefont {W.~P.}\ \bibnamefont {Pratt}}, \ and\ \bibinfo {author} {\bibfnamefont {N.~O.}\ \bibnamefont {Birge}},\ }\bibfield  {title} {\enquote {\bibinfo {title} {Observation of standard spin-switch effects in ferromagnet/superconductor/ferromagnet trilayers with a strong ferromagnet},}\ }\href {\doibase 10.1103/PhysRevB.74.220507} {\bibfield  {journal} {\bibinfo  {journal} {Phys. Rev. B}\ }\textbf {\bibinfo {volume} {74}},\ \bibinfo {pages} {220507} (\bibinfo {year} {2006})}\BibitemShut {NoStop}%
\bibitem [{\citenamefont {Leksin}\ \emph {et~al.}(2012)\citenamefont {Leksin}, \citenamefont {Garif'yanov}, \citenamefont {Garifullin}, \citenamefont {Fominov}, \citenamefont {Schumann}, \citenamefont {Krupskaya}, \citenamefont {Kataev}, \citenamefont {Schmidt},\ and\ \citenamefont {B\"uchner}}]{Leksin2012}%
  \BibitemOpen
  \bibfield  {author} {\bibinfo {author} {\bibfnamefont {P.~V.}\ \bibnamefont {Leksin}}, \bibinfo {author} {\bibfnamefont {N.~N.}\ \bibnamefont {Garif'yanov}}, \bibinfo {author} {\bibfnamefont {I.~A.}\ \bibnamefont {Garifullin}}, \bibinfo {author} {\bibfnamefont {Y.~V.}\ \bibnamefont {Fominov}}, \bibinfo {author} {\bibfnamefont {J.}~\bibnamefont {Schumann}}, \bibinfo {author} {\bibfnamefont {Y.}~\bibnamefont {Krupskaya}}, \bibinfo {author} {\bibfnamefont {V.}~\bibnamefont {Kataev}}, \bibinfo {author} {\bibfnamefont {O.~G.}\ \bibnamefont {Schmidt}}, \ and\ \bibinfo {author} {\bibfnamefont {B.}~\bibnamefont {B\"uchner}},\ }\bibfield  {title} {\enquote {\bibinfo {title} {Evidence for triplet superconductivity in a superconductor-ferromagnet spin valve},}\ }\href {\doibase 10.1103/PhysRevLett.109.057005} {\bibfield  {journal} {\bibinfo  {journal} {Phys. Rev. Lett.}\ }\textbf {\bibinfo {volume} {109}},\ \bibinfo {pages} {057005} (\bibinfo {year} {2012})}\BibitemShut {NoStop}%
\bibitem [{\citenamefont {Wang}\ \emph {et~al.}(2014)\citenamefont {Wang}, \citenamefont {Di~Bernardo}, \citenamefont {Banerjee}, \citenamefont {Wells}, \citenamefont {Bergeret}, \citenamefont {Blamire},\ and\ \citenamefont {Robinson}}]{Wang2014}%
  \BibitemOpen
  \bibfield  {author} {\bibinfo {author} {\bibfnamefont {X.~L.}\ \bibnamefont {Wang}}, \bibinfo {author} {\bibfnamefont {A.}~\bibnamefont {Di~Bernardo}}, \bibinfo {author} {\bibfnamefont {N.}~\bibnamefont {Banerjee}}, \bibinfo {author} {\bibfnamefont {A.}~\bibnamefont {Wells}}, \bibinfo {author} {\bibfnamefont {F.~S.}\ \bibnamefont {Bergeret}}, \bibinfo {author} {\bibfnamefont {M.~G.}\ \bibnamefont {Blamire}}, \ and\ \bibinfo {author} {\bibfnamefont {J.~W.~A.}\ \bibnamefont {Robinson}},\ }\bibfield  {title} {\enquote {\bibinfo {title} {Giant triplet proximity effect in superconducting pseudo spin valves with engineered anisotropy},}\ }\href {\doibase 10.1103/PhysRevB.89.140508} {\bibfield  {journal} {\bibinfo  {journal} {Phys. Rev. B}\ }\textbf {\bibinfo {volume} {89}},\ \bibinfo {pages} {140508} (\bibinfo {year} {2014})}\BibitemShut {NoStop}%
\bibitem [{\citenamefont {Zdravkov}\ \emph {et~al.}(2013)\citenamefont {Zdravkov}, \citenamefont {Kehrle}, \citenamefont {Obermeier}, \citenamefont {Lenk}, \citenamefont {Krug~von Nidda}, \citenamefont {M\"uller}, \citenamefont {Kupriyanov}, \citenamefont {Sidorenko}, \citenamefont {Horn}, \citenamefont {Tidecks},\ and\ \citenamefont {Tagirov}}]{Zdravkov2013}%
  \BibitemOpen
  \bibfield  {author} {\bibinfo {author} {\bibfnamefont {V.~I.}\ \bibnamefont {Zdravkov}}, \bibinfo {author} {\bibfnamefont {J.}~\bibnamefont {Kehrle}}, \bibinfo {author} {\bibfnamefont {G.}~\bibnamefont {Obermeier}}, \bibinfo {author} {\bibfnamefont {D.}~\bibnamefont {Lenk}}, \bibinfo {author} {\bibfnamefont {H.-A.}\ \bibnamefont {Krug~von Nidda}}, \bibinfo {author} {\bibfnamefont {C.}~\bibnamefont {M\"uller}}, \bibinfo {author} {\bibfnamefont {M.~Y.}\ \bibnamefont {Kupriyanov}}, \bibinfo {author} {\bibfnamefont {A.~S.}\ \bibnamefont {Sidorenko}}, \bibinfo {author} {\bibfnamefont {S.}~\bibnamefont {Horn}}, \bibinfo {author} {\bibfnamefont {R.}~\bibnamefont {Tidecks}}, \ and\ \bibinfo {author} {\bibfnamefont {L.~R.}\ \bibnamefont {Tagirov}},\ }\bibfield  {title} {\enquote {\bibinfo {title} {Experimental observation of the triplet spin-valve effect in a superconductor-ferromagnet heterostructure},}\ }\href {\doibase 10.1103/PhysRevB.87.144507} {\bibfield  {journal} {\bibinfo  {journal} {Phys. Rev. B}\ }\textbf
  {\bibinfo {volume} {87}},\ \bibinfo {pages} {144507} (\bibinfo {year} {2013})}\BibitemShut {NoStop}%
\bibitem [{\citenamefont {Banerjee}\ \emph {et~al.}(2014)\citenamefont {Banerjee}, \citenamefont {Smiet}, \citenamefont {Smits}, \citenamefont {Ozaeta}, \citenamefont {Bergeret}, \citenamefont {Blamire},\ and\ \citenamefont {Robinson}}]{Banerjee2014}%
  \BibitemOpen
  \bibfield  {author} {\bibinfo {author} {\bibfnamefont {N.}~\bibnamefont {Banerjee}}, \bibinfo {author} {\bibfnamefont {C.}~\bibnamefont {Smiet}}, \bibinfo {author} {\bibfnamefont {R.}~\bibnamefont {Smits}}, \bibinfo {author} {\bibfnamefont {A.}~\bibnamefont {Ozaeta}}, \bibinfo {author} {\bibfnamefont {F.}~\bibnamefont {Bergeret}}, \bibinfo {author} {\bibfnamefont {M.}~\bibnamefont {Blamire}}, \ and\ \bibinfo {author} {\bibfnamefont {J.}~\bibnamefont {Robinson}},\ }\bibfield  {title} {\enquote {\bibinfo {title} {Evidence for spin selectivity of triplet pairs in superconducting spin valves},}\ }\href {\doibase 10.1038/ncomms4048} {\bibfield  {journal} {\bibinfo  {journal} {Nature Communications}\ }\textbf {\bibinfo {volume} {5}},\ \bibinfo {pages} {3048} (\bibinfo {year} {2014})}\BibitemShut {NoStop}%
\bibitem [{\citenamefont {Bergeret}, \citenamefont {Volkov},\ and\ \citenamefont {Efetov}(2001)}]{Bergeret2001}%
  \BibitemOpen
  \bibfield  {author} {\bibinfo {author} {\bibfnamefont {F.~S.}\ \bibnamefont {Bergeret}}, \bibinfo {author} {\bibfnamefont {A.~F.}\ \bibnamefont {Volkov}}, \ and\ \bibinfo {author} {\bibfnamefont {K.~B.}\ \bibnamefont {Efetov}},\ }\bibfield  {title} {\enquote {\bibinfo {title} {Long-range proximity effects in superconductor-ferromagnet structures},}\ }\href {\doibase 10.1103/PhysRevLett.86.4096} {\bibfield  {journal} {\bibinfo  {journal} {Phys. Rev. Lett.}\ }\textbf {\bibinfo {volume} {86}},\ \bibinfo {pages} {4096--4099} (\bibinfo {year} {2001})}\BibitemShut {NoStop}%
\bibitem [{\citenamefont {Khaire}\ \emph {et~al.}(2010)\citenamefont {Khaire}, \citenamefont {Khasawneh}, \citenamefont {Pratt},\ and\ \citenamefont {Birge}}]{Khaire2010}%
  \BibitemOpen
  \bibfield  {author} {\bibinfo {author} {\bibfnamefont {T.~S.}\ \bibnamefont {Khaire}}, \bibinfo {author} {\bibfnamefont {M.~A.}\ \bibnamefont {Khasawneh}}, \bibinfo {author} {\bibfnamefont {W.~P.}\ \bibnamefont {Pratt}}, \ and\ \bibinfo {author} {\bibfnamefont {N.~O.}\ \bibnamefont {Birge}},\ }\bibfield  {title} {\enquote {\bibinfo {title} {Observation of spin-triplet superconductivity in co-based josephson junctions},}\ }\href {\doibase 10.1103/PhysRevLett.104.137002} {\bibfield  {journal} {\bibinfo  {journal} {Phys. Rev. Lett.}\ }\textbf {\bibinfo {volume} {104}},\ \bibinfo {pages} {137002} (\bibinfo {year} {2010})}\BibitemShut {NoStop}%
\bibitem [{\citenamefont {Robinson}, \citenamefont {Witt},\ and\ \citenamefont {Blamire}(2010)}]{Robinson2010}%
  \BibitemOpen
  \bibfield  {author} {\bibinfo {author} {\bibfnamefont {J.~W.~A.}\ \bibnamefont {Robinson}}, \bibinfo {author} {\bibfnamefont {J.~D.~S.}\ \bibnamefont {Witt}}, \ and\ \bibinfo {author} {\bibfnamefont {M.~G.}\ \bibnamefont {Blamire}},\ }\bibfield  {title} {\enquote {\bibinfo {title} {Controlled injection of spin-triplet supercurrents into a strong ferromagnet},}\ }\href {\doibase 10.1126/science.1189246} {\bibfield  {journal} {\bibinfo  {journal} {Science}\ }\textbf {\bibinfo {volume} {329}},\ \bibinfo {pages} {59--61} (\bibinfo {year} {2010})}\BibitemShut {NoStop}%
\bibitem [{\citenamefont {Banerjee}\ \emph {et~al.}(2018)\citenamefont {Banerjee}, \citenamefont {Ouassou}, \citenamefont {Zhu}, \citenamefont {Stelmashenko}, \citenamefont {Linder},\ and\ \citenamefont {Blamire}}]{Banerjee2018}%
  \BibitemOpen
  \bibfield  {author} {\bibinfo {author} {\bibfnamefont {N.}~\bibnamefont {Banerjee}}, \bibinfo {author} {\bibfnamefont {J.~A.}\ \bibnamefont {Ouassou}}, \bibinfo {author} {\bibfnamefont {Y.}~\bibnamefont {Zhu}}, \bibinfo {author} {\bibfnamefont {N.~A.}\ \bibnamefont {Stelmashenko}}, \bibinfo {author} {\bibfnamefont {J.}~\bibnamefont {Linder}}, \ and\ \bibinfo {author} {\bibfnamefont {M.~G.}\ \bibnamefont {Blamire}},\ }\bibfield  {title} {\enquote {\bibinfo {title} {Controlling the superconducting transition by spin-orbit coupling},}\ }\href {\doibase 10.1103/PhysRevB.97.184521} {\bibfield  {journal} {\bibinfo  {journal} {Phys. Rev. B}\ }\textbf {\bibinfo {volume} {97}},\ \bibinfo {pages} {184521} (\bibinfo {year} {2018})}\BibitemShut {NoStop}%
\bibitem [{\citenamefont {Bergeret}\ and\ \citenamefont {Tokatly}(2013)}]{Bergeret2013}%
  \BibitemOpen
  \bibfield  {author} {\bibinfo {author} {\bibfnamefont {F.~S.}\ \bibnamefont {Bergeret}}\ and\ \bibinfo {author} {\bibfnamefont {I.~V.}\ \bibnamefont {Tokatly}},\ }\bibfield  {title} {\enquote {\bibinfo {title} {Singlet-triplet conversion and the long-range proximity effect in superconductor-ferromagnet structures with generic spin dependent fields},}\ }\href {\doibase 10.1103/PhysRevLett.110.117003} {\bibfield  {journal} {\bibinfo  {journal} {Phys. Rev. Lett.}\ }\textbf {\bibinfo {volume} {110}},\ \bibinfo {pages} {117003} (\bibinfo {year} {2013})}\BibitemShut {NoStop}%
\bibitem [{\citenamefont {Bergeret}\ and\ \citenamefont {Tokatly}(2014)}]{Bergeret2014}%
  \BibitemOpen
  \bibfield  {author} {\bibinfo {author} {\bibfnamefont {F.~S.}\ \bibnamefont {Bergeret}}\ and\ \bibinfo {author} {\bibfnamefont {I.~V.}\ \bibnamefont {Tokatly}},\ }\href {\doibase 10.1103/PhysRevB.89.134517} {\bibfield  {journal} {\bibinfo  {journal} {Physical Review B}\ }\textbf {\bibinfo {volume} {89}},\ \bibinfo {pages} {134517} (\bibinfo {year} {2014})}\BibitemShut {NoStop}%
\bibitem [{\citenamefont {Amundsen}\ \emph {et~al.}(2022)\citenamefont {Amundsen}, \citenamefont {Linder}, \citenamefont {Robinson}, \citenamefont {Žutić},\ and\ \citenamefont {Banerjee}}]{amundsen2022}%
  \BibitemOpen
  \bibfield  {author} {\bibinfo {author} {\bibfnamefont {M.}~\bibnamefont {Amundsen}}, \bibinfo {author} {\bibfnamefont {J.}~\bibnamefont {Linder}}, \bibinfo {author} {\bibfnamefont {J.~W.~A.}\ \bibnamefont {Robinson}}, \bibinfo {author} {\bibfnamefont {I.}~\bibnamefont {Žutić}}, \ and\ \bibinfo {author} {\bibfnamefont {N.}~\bibnamefont {Banerjee}},\ }\href@noop {} {\enquote {\bibinfo {title} {Colloquium: Spin-orbit effects in superconducting hybrid structures},}\ } (\bibinfo {year} {2022}),\ \Eprint {http://arxiv.org/abs/2210.03549} {arXiv:2210.03549 [cond-mat.supr-con]} \BibitemShut {NoStop}%
\bibitem [{\citenamefont {Jacobsen}, \citenamefont {Ouassou},\ and\ \citenamefont {Linder}(2015)}]{Jacobsen2015b}%
  \BibitemOpen
  \bibfield  {author} {\bibinfo {author} {\bibfnamefont {S.~H.}\ \bibnamefont {Jacobsen}}, \bibinfo {author} {\bibfnamefont {J.~A.}\ \bibnamefont {Ouassou}}, \ and\ \bibinfo {author} {\bibfnamefont {J.}~\bibnamefont {Linder}},\ }\bibfield  {title} {\enquote {\bibinfo {title} {{Critical Temperature and Tunneling Spectroscopy of Superconductor-Ferromagnet Hybrids with Intrinsic Rashba–Dresselhaus Spin-Orbit Coupling}},}\ }\href {\doibase 10.1103/PhysRevB.92.024510} {\bibfield  {journal} {\bibinfo  {journal} {Physical Review B}\ }\textbf {\bibinfo {volume} {92}},\ \bibinfo {pages} {024510} (\bibinfo {year} {2015})}\BibitemShut {NoStop}%
\bibitem [{\citenamefont {Ouassou}\ \emph {et~al.}(2016)\citenamefont {Ouassou}, \citenamefont {Di~Bernardo}, \citenamefont {Robinson},\ and\ \citenamefont {Linder}}]{Ouassou2016}%
  \BibitemOpen
  \bibfield  {author} {\bibinfo {author} {\bibfnamefont {J.}~\bibnamefont {Ouassou}}, \bibinfo {author} {\bibfnamefont {A.}~\bibnamefont {Di~Bernardo}}, \bibinfo {author} {\bibfnamefont {J.}~\bibnamefont {Robinson}}, \ and\ \bibinfo {author} {\bibfnamefont {J.}~\bibnamefont {Linder}},\ }\bibfield  {title} {\enquote {\bibinfo {title} {Electric control of superconducting transition through a spin-orbit coupled interface},}\ }\href {\doibase 10.1038/srep29312} {\bibfield  {journal} {\bibinfo  {journal} {Scientific Reports}\ }\textbf {\bibinfo {volume} {6}},\ \bibinfo {pages} {29312} (\bibinfo {year} {2016})}\BibitemShut {NoStop}%
\bibitem [{\citenamefont {Salamone}\ \emph {et~al.}(2021)\citenamefont {Salamone}, \citenamefont {Svendsen}, \citenamefont {Amundsen},\ and\ \citenamefont {Jacobsen}}]{Salamone2021}%
  \BibitemOpen
  \bibfield  {author} {\bibinfo {author} {\bibfnamefont {T.}~\bibnamefont {Salamone}}, \bibinfo {author} {\bibfnamefont {M.~B.~M.}\ \bibnamefont {Svendsen}}, \bibinfo {author} {\bibfnamefont {M.}~\bibnamefont {Amundsen}}, \ and\ \bibinfo {author} {\bibfnamefont {S.}~\bibnamefont {Jacobsen}},\ }\bibfield  {title} {\enquote {\bibinfo {title} {Curvature-induced long-range supercurrents in diffusive superconductor-ferromagnet-superconductor {J}osephson junctions with a dynamic $0\text{\ensuremath{-}}\ensuremath{\pi}$ transition},}\ }\href {\doibase 10.1103/PhysRevB.104.L060505} {\bibfield  {journal} {\bibinfo  {journal} {Phys. Rev. B}\ }\textbf {\bibinfo {volume} {104}},\ \bibinfo {pages} {L060505} (\bibinfo {year} {2021})}\BibitemShut {NoStop}%
\bibitem [{\citenamefont {Salamone}\ \emph {et~al.}(2022)\citenamefont {Salamone}, \citenamefont {Hugdal}, \citenamefont {Amundsen},\ and\ \citenamefont {Jacobsen}}]{Salamone2022}%
  \BibitemOpen
  \bibfield  {author} {\bibinfo {author} {\bibfnamefont {T.}~\bibnamefont {Salamone}}, \bibinfo {author} {\bibfnamefont {H.~G.}\ \bibnamefont {Hugdal}}, \bibinfo {author} {\bibfnamefont {M.}~\bibnamefont {Amundsen}}, \ and\ \bibinfo {author} {\bibfnamefont {S.~H.}\ \bibnamefont {Jacobsen}},\ }\bibfield  {title} {\enquote {\bibinfo {title} {Curvature control of the superconducting proximity effect in diffusive ferromagnetic nanowires},}\ }\href {\doibase 10.1103/PhysRevB.105.134511} {\bibfield  {journal} {\bibinfo  {journal} {Phys. Rev. B}\ }\textbf {\bibinfo {volume} {105}},\ \bibinfo {pages} {134511} (\bibinfo {year} {2022})}\BibitemShut {NoStop}%
\bibitem [{\citenamefont {Skarpeid}\ \emph {et~al.}(2024)\citenamefont {Skarpeid}, \citenamefont {Hugdal}, \citenamefont {Salamone}, \citenamefont {Amundsen},\ and\ \citenamefont {Jacobsen}}]{Skarpeid2024}%
  \BibitemOpen
  \bibfield  {author} {\bibinfo {author} {\bibfnamefont {A.~J.}\ \bibnamefont {Skarpeid}}, \bibinfo {author} {\bibfnamefont {H.~G.}\ \bibnamefont {Hugdal}}, \bibinfo {author} {\bibfnamefont {T.}~\bibnamefont {Salamone}}, \bibinfo {author} {\bibfnamefont {M.}~\bibnamefont {Amundsen}}, \ and\ \bibinfo {author} {\bibfnamefont {S.~H.}\ \bibnamefont {Jacobsen}},\ }\bibfield  {title} {\enquote {\bibinfo {title} {Non-constant geometric curvature for tailored spin-orbit coupling and chirality in superconductor-magnet heterostructures},}\ }\href {http://iopscience.iop.org/article/10.1088/1361-648X/ad2e23} {\bibfield  {journal} {\bibinfo  {journal} {Journal of Physics: Condensed Matter}\ }\textbf {\bibinfo {volume} {36}},\ \bibinfo {pages} {235302} (\bibinfo {year} {2024})}\BibitemShut {NoStop}%
\bibitem [{\citenamefont {Salamone}\ \emph {et~al.}(2024)\citenamefont {Salamone}, \citenamefont {Skj\ae{}rpe}, \citenamefont {Hugdal}, \citenamefont {Amundsen},\ and\ \citenamefont {Jacobsen}}]{Salamone2024}%
  \BibitemOpen
  \bibfield  {author} {\bibinfo {author} {\bibfnamefont {T.}~\bibnamefont {Salamone}}, \bibinfo {author} {\bibfnamefont {M.}~\bibnamefont {Skj\ae{}rpe}}, \bibinfo {author} {\bibfnamefont {H.~G.}\ \bibnamefont {Hugdal}}, \bibinfo {author} {\bibfnamefont {M.}~\bibnamefont {Amundsen}}, \ and\ \bibinfo {author} {\bibfnamefont {S.~H.}\ \bibnamefont {Jacobsen}},\ }\bibfield  {title} {\enquote {\bibinfo {title} {Interface probe for antiferromagnets using geometric curvature},}\ }\href {\doibase 10.1103/PhysRevB.109.094508} {\bibfield  {journal} {\bibinfo  {journal} {Phys. Rev. B}\ }\textbf {\bibinfo {volume} {109}},\ \bibinfo {pages} {094508} (\bibinfo {year} {2024})}\BibitemShut {NoStop}%
\bibitem [{\citenamefont {Gentile}\ \emph {et~al.}(2022)\citenamefont {Gentile}, \citenamefont {Cuoco}, \citenamefont {Volkov}, \citenamefont {Ying}, \citenamefont {Vera-Marun}, \citenamefont {Makarov},\ and\ \citenamefont {Ortix}}]{Gentile2022}%
  \BibitemOpen
  \bibfield  {author} {\bibinfo {author} {\bibfnamefont {P.}~\bibnamefont {Gentile}}, \bibinfo {author} {\bibfnamefont {M.}~\bibnamefont {Cuoco}}, \bibinfo {author} {\bibfnamefont {O.~M.}\ \bibnamefont {Volkov}}, \bibinfo {author} {\bibfnamefont {Z.-J.}\ \bibnamefont {Ying}}, \bibinfo {author} {\bibfnamefont {I.~J.}\ \bibnamefont {Vera-Marun}}, \bibinfo {author} {\bibfnamefont {D.}~\bibnamefont {Makarov}}, \ and\ \bibinfo {author} {\bibfnamefont {C.}~\bibnamefont {Ortix}},\ }\bibfield  {title} {\enquote {\bibinfo {title} {Electronic materials with nanoscale curved geometries},}\ }\href {\doibase 10.1038/s41928-022-00820-z} {\bibfield  {journal} {\bibinfo  {journal} {Nature Electronics}\ }\textbf {\bibinfo {volume} {5}},\ \bibinfo {pages} {551--563} (\bibinfo {year} {2022})}\BibitemShut {NoStop}%
\bibitem [{\citenamefont {Streubel}\ \emph {et~al.}(2016)\citenamefont {Streubel}, \citenamefont {Fischer}, \citenamefont {Kronast}, \citenamefont {Kravchuk}, \citenamefont {Sheka}, \citenamefont {Gaididei}, \citenamefont {Schmidt},\ and\ \citenamefont {Makarov}}]{Streubel2016}%
  \BibitemOpen
  \bibfield  {author} {\bibinfo {author} {\bibfnamefont {R.}~\bibnamefont {Streubel}}, \bibinfo {author} {\bibfnamefont {P.}~\bibnamefont {Fischer}}, \bibinfo {author} {\bibfnamefont {F.}~\bibnamefont {Kronast}}, \bibinfo {author} {\bibfnamefont {V.~P.}\ \bibnamefont {Kravchuk}}, \bibinfo {author} {\bibfnamefont {D.~D.}\ \bibnamefont {Sheka}}, \bibinfo {author} {\bibfnamefont {Y.}~\bibnamefont {Gaididei}}, \bibinfo {author} {\bibfnamefont {O.~G.}\ \bibnamefont {Schmidt}}, \ and\ \bibinfo {author} {\bibfnamefont {D.}~\bibnamefont {Makarov}},\ }\bibfield  {title} {\enquote {\bibinfo {title} {Magnetism in curved geometries},}\ }\href {\doibase 10.1088/0022-3727/49/36/363001} {\bibfield  {journal} {\bibinfo  {journal} {Journal of Physics D: Applied Physics}\ }\textbf {\bibinfo {volume} {49}},\ \bibinfo {pages} {363001} (\bibinfo {year} {2016})}\BibitemShut {NoStop}%
\bibitem [{\citenamefont {Streubel}, \citenamefont {Tsymbal},\ and\ \citenamefont {Fischer}(2021)}]{Streubel2021}%
  \BibitemOpen
  \bibfield  {author} {\bibinfo {author} {\bibfnamefont {R.}~\bibnamefont {Streubel}}, \bibinfo {author} {\bibfnamefont {E.~Y.}\ \bibnamefont {Tsymbal}}, \ and\ \bibinfo {author} {\bibfnamefont {P.}~\bibnamefont {Fischer}},\ }\bibfield  {title} {\enquote {\bibinfo {title} {Magnetism in curved geometries},}\ }\href {\doibase https://doi.org/10.1063/5.0054025} {\bibfield  {journal} {\bibinfo  {journal} {Journal of Applied Physics}\ }\textbf {\bibinfo {volume} {129}},\ \bibinfo {pages} {210902} (\bibinfo {year} {2021})}\BibitemShut {NoStop}%
\bibitem [{\citenamefont {Gentile}, \citenamefont {Cuoco},\ and\ \citenamefont {Ortix}(2013)}]{Gentile2013}%
  \BibitemOpen
  \bibfield  {author} {\bibinfo {author} {\bibfnamefont {P.}~\bibnamefont {Gentile}}, \bibinfo {author} {\bibfnamefont {M.}~\bibnamefont {Cuoco}}, \ and\ \bibinfo {author} {\bibfnamefont {C.}~\bibnamefont {Ortix}},\ }\bibfield  {title} {\enquote {\bibinfo {title} {{Curvature-induced rashba spin-orbit interaction in strain-driven nanostructures}},}\ }\href {\doibase 10.1142/S201032471340002X} {\bibfield  {journal} {\bibinfo  {journal} {Spin}\ }\textbf {\bibinfo {volume} {3}},\ \bibinfo {pages} {1340002} (\bibinfo {year} {2013})}\BibitemShut {NoStop}%
\bibitem [{\citenamefont {Sheka}\ \emph {et~al.}(2022)\citenamefont {Sheka}, \citenamefont {Pylypovskyi}, \citenamefont {Volkov}, \citenamefont {Yershov}, \citenamefont {Kravchuk},\ and\ \citenamefont {Makarov}}]{Sheka2022}%
  \BibitemOpen
  \bibfield  {author} {\bibinfo {author} {\bibfnamefont {D.~D.}\ \bibnamefont {Sheka}}, \bibinfo {author} {\bibfnamefont {O.~V.}\ \bibnamefont {Pylypovskyi}}, \bibinfo {author} {\bibfnamefont {O.~M.}\ \bibnamefont {Volkov}}, \bibinfo {author} {\bibfnamefont {K.~V.}\ \bibnamefont {Yershov}}, \bibinfo {author} {\bibfnamefont {V.~P.}\ \bibnamefont {Kravchuk}}, \ and\ \bibinfo {author} {\bibfnamefont {D.}~\bibnamefont {Makarov}},\ }\bibfield  {title} {\enquote {\bibinfo {title} {Fundamentals of {{Curvilinear Ferromagnetism}}: {{Statics}} and {{Dynamics}} of {{Geometrically Curved Wires}} and {{Narrow Ribbons}}},}\ }\href {\doibase 10.1002/smll.202105219} {\bibfield  {journal} {\bibinfo  {journal} {Small}\ }\textbf {\bibinfo {volume} {18}},\ \bibinfo {pages} {2105219} (\bibinfo {year} {2022})}\BibitemShut {NoStop}%
\bibitem [{\citenamefont {Makarov}\ \emph {et~al.}(2022)\citenamefont {Makarov}, \citenamefont {Volkov}, \citenamefont {Kákay}, \citenamefont {Pylypovskyi}, \citenamefont {Budinská},\ and\ \citenamefont {Dobrovolskiy}}]{Makarov2022}%
  \BibitemOpen
  \bibfield  {author} {\bibinfo {author} {\bibfnamefont {D.}~\bibnamefont {Makarov}}, \bibinfo {author} {\bibfnamefont {O.~M.}\ \bibnamefont {Volkov}}, \bibinfo {author} {\bibfnamefont {A.}~\bibnamefont {Kákay}}, \bibinfo {author} {\bibfnamefont {O.~V.}\ \bibnamefont {Pylypovskyi}}, \bibinfo {author} {\bibfnamefont {B.}~\bibnamefont {Budinská}}, \ and\ \bibinfo {author} {\bibfnamefont {O.~V.}\ \bibnamefont {Dobrovolskiy}},\ }\bibfield  {title} {\enquote {\bibinfo {title} {New dimension in magnetism and superconductivity: 3d and curvilinear nanoarchitectures},}\ }\href {\doibase https://doi.org/10.1002/adma.202101758} {\bibfield  {journal} {\bibinfo  {journal} {Advanced Materials}\ }\textbf {\bibinfo {volume} {34}},\ \bibinfo {pages} {2101758} (\bibinfo {year} {2022})}\BibitemShut {NoStop}%
\bibitem [{\citenamefont {Volkov}\ \emph {et~al.}(2019{\natexlab{a}})\citenamefont {Volkov}, \citenamefont {K\'akay}, \citenamefont {Kronast}, \citenamefont {M\"onch}, \citenamefont {Mawass}, \citenamefont {Fassbender},\ and\ \citenamefont {Makarov}}]{Volkov2019}%
  \BibitemOpen
  \bibfield  {author} {\bibinfo {author} {\bibfnamefont {O.~M.}\ \bibnamefont {Volkov}}, \bibinfo {author} {\bibfnamefont {A.}~\bibnamefont {K\'akay}}, \bibinfo {author} {\bibfnamefont {F.}~\bibnamefont {Kronast}}, \bibinfo {author} {\bibfnamefont {I.}~\bibnamefont {M\"onch}}, \bibinfo {author} {\bibfnamefont {M.-A.}\ \bibnamefont {Mawass}}, \bibinfo {author} {\bibfnamefont {J.}~\bibnamefont {Fassbender}}, \ and\ \bibinfo {author} {\bibfnamefont {D.}~\bibnamefont {Makarov}},\ }\bibfield  {title} {\enquote {\bibinfo {title} {Experimental observation of exchange-driven chiral effects in curvilinear magnetism},}\ }\href {\doibase 10.1103/PhysRevLett.123.077201} {\bibfield  {journal} {\bibinfo  {journal} {Phys. Rev. Lett.}\ }\textbf {\bibinfo {volume} {123}},\ \bibinfo {pages} {077201} (\bibinfo {year} {2019}{\natexlab{a}})}\BibitemShut {NoStop}%
\bibitem [{\citenamefont {Sahoo}\ \emph {et~al.}(2018)\citenamefont {Sahoo}, \citenamefont {Mondal}, \citenamefont {Williams}, \citenamefont {May}, \citenamefont {Ladak},\ and\ \citenamefont {Barman}}]{Sahoo2018}%
  \BibitemOpen
  \bibfield  {author} {\bibinfo {author} {\bibfnamefont {S.}~\bibnamefont {Sahoo}}, \bibinfo {author} {\bibfnamefont {S.}~\bibnamefont {Mondal}}, \bibinfo {author} {\bibfnamefont {G.}~\bibnamefont {Williams}}, \bibinfo {author} {\bibfnamefont {A.}~\bibnamefont {May}}, \bibinfo {author} {\bibfnamefont {S.}~\bibnamefont {Ladak}}, \ and\ \bibinfo {author} {\bibfnamefont {A.}~\bibnamefont {Barman}},\ }\bibfield  {title} {\enquote {\bibinfo {title} {Ultrafast magnetization dynamics in a nanoscale three-dimensional cobalt tetrapod structure},}\ }\href {\doibase 10.1039/C7NR07843A} {\bibfield  {journal} {\bibinfo  {journal} {Nanoscale}\ }\textbf {\bibinfo {volume} {10}},\ \bibinfo {pages} {9981--9986} (\bibinfo {year} {2018})}\BibitemShut {NoStop}%
\bibitem [{\citenamefont {Dobrovolskiy}\ \emph {et~al.}(2021)\citenamefont {Dobrovolskiy}, \citenamefont {Vovk}, \citenamefont {Bondarenko}, \citenamefont {Bunyaev}, \citenamefont {Lamb-Camarena}, \citenamefont {Zenbaa}, \citenamefont {Sachser}, \citenamefont {Barth}, \citenamefont {Guslienko}, \citenamefont {Chumak}, \citenamefont {Huth},\ and\ \citenamefont {Kakazei}}]{Dobrovolskiy2021}%
  \BibitemOpen
  \bibfield  {author} {\bibinfo {author} {\bibfnamefont {O.~V.}\ \bibnamefont {Dobrovolskiy}}, \bibinfo {author} {\bibfnamefont {N.~R.}\ \bibnamefont {Vovk}}, \bibinfo {author} {\bibfnamefont {A.~V.}\ \bibnamefont {Bondarenko}}, \bibinfo {author} {\bibfnamefont {S.~A.}\ \bibnamefont {Bunyaev}}, \bibinfo {author} {\bibfnamefont {S.}~\bibnamefont {Lamb-Camarena}}, \bibinfo {author} {\bibfnamefont {N.}~\bibnamefont {Zenbaa}}, \bibinfo {author} {\bibfnamefont {R.}~\bibnamefont {Sachser}}, \bibinfo {author} {\bibfnamefont {S.}~\bibnamefont {Barth}}, \bibinfo {author} {\bibfnamefont {K.~Y.}\ \bibnamefont {Guslienko}}, \bibinfo {author} {\bibfnamefont {A.~V.}\ \bibnamefont {Chumak}}, \bibinfo {author} {\bibfnamefont {M.}~\bibnamefont {Huth}}, \ and\ \bibinfo {author} {\bibfnamefont {G.~N.}\ \bibnamefont {Kakazei}},\ }\bibfield  {title} {\enquote {\bibinfo {title} {{Spin-wave eigenmodes in direct-write 3D nanovolcanoes}},}\ }\href {\doibase 10.1063/5.0044325} {\bibfield  {journal} {\bibinfo  {journal} {Applied
  Physics Letters}\ }\textbf {\bibinfo {volume} {118}},\ \bibinfo {pages} {132405} (\bibinfo {year} {2021})}\BibitemShut {NoStop}%
\bibitem [{\citenamefont {Sanz-Hern{\'a}ndez}\ \emph {et~al.}(2020)\citenamefont {Sanz-Hern{\'a}ndez}, \citenamefont {Hierro-Rodriguez}, \citenamefont {Donnelly}, \citenamefont {Pablo-Navarro}, \citenamefont {Sorrentino}, \citenamefont {Pereiro}, \citenamefont {Mag{\'e}n}, \citenamefont {McVitie}, \citenamefont {de~Teresa}, \citenamefont {Ferrer}, \citenamefont {Fischer},\ and\ \citenamefont {Fern{\'a}ndez-Pacheco}}]{Sanz-Hernandez2020}%
  \BibitemOpen
  \bibfield  {author} {\bibinfo {author} {\bibfnamefont {D.}~\bibnamefont {Sanz-Hern{\'a}ndez}}, \bibinfo {author} {\bibfnamefont {A.}~\bibnamefont {Hierro-Rodriguez}}, \bibinfo {author} {\bibfnamefont {C.}~\bibnamefont {Donnelly}}, \bibinfo {author} {\bibfnamefont {J.}~\bibnamefont {Pablo-Navarro}}, \bibinfo {author} {\bibfnamefont {A.}~\bibnamefont {Sorrentino}}, \bibinfo {author} {\bibfnamefont {E.}~\bibnamefont {Pereiro}}, \bibinfo {author} {\bibfnamefont {C.}~\bibnamefont {Mag{\'e}n}}, \bibinfo {author} {\bibfnamefont {S.}~\bibnamefont {McVitie}}, \bibinfo {author} {\bibfnamefont {J.~M.}\ \bibnamefont {de~Teresa}}, \bibinfo {author} {\bibfnamefont {S.}~\bibnamefont {Ferrer}}, \bibinfo {author} {\bibfnamefont {P.}~\bibnamefont {Fischer}}, \ and\ \bibinfo {author} {\bibfnamefont {A.}~\bibnamefont {Fern{\'a}ndez-Pacheco}},\ }\bibfield  {title} {\enquote {\bibinfo {title} {Artificial double-helix for geometrical control of magnetic chirality},}\ }\href {\doibase 10.1021/acsnano.0c00720} {\bibfield  {journal}
  {\bibinfo  {journal} {ACS Nano}\ }\textbf {\bibinfo {volume} {14}},\ \bibinfo {pages} {8084--8092} (\bibinfo {year} {2020})}\BibitemShut {NoStop}%
\bibitem [{\citenamefont {Skoric}\ \emph {et~al.}(2020)\citenamefont {Skoric}, \citenamefont {Sanz-Hern{\'a}ndez}, \citenamefont {Meng}, \citenamefont {Donnelly}, \citenamefont {Merino-Aceituno},\ and\ \citenamefont {Fern{\'a}ndez-Pacheco}}]{Skoric2020}%
  \BibitemOpen
  \bibfield  {author} {\bibinfo {author} {\bibfnamefont {L.}~\bibnamefont {Skoric}}, \bibinfo {author} {\bibfnamefont {D.}~\bibnamefont {Sanz-Hern{\'a}ndez}}, \bibinfo {author} {\bibfnamefont {F.}~\bibnamefont {Meng}}, \bibinfo {author} {\bibfnamefont {C.}~\bibnamefont {Donnelly}}, \bibinfo {author} {\bibfnamefont {S.}~\bibnamefont {Merino-Aceituno}}, \ and\ \bibinfo {author} {\bibfnamefont {A.}~\bibnamefont {Fern{\'a}ndez-Pacheco}},\ }\bibfield  {title} {\enquote {\bibinfo {title} {Layer-by-layer growth of complex-shaped three-dimensional nanostructures with focused electron beams},}\ }\href {\doibase 10.1021/acs.nanolett.9b03565} {\bibfield  {journal} {\bibinfo  {journal} {Nano Letters}\ }\textbf {\bibinfo {volume} {20}},\ \bibinfo {pages} {184--191} (\bibinfo {year} {2020})}\BibitemShut {NoStop}%
\bibitem [{\citenamefont {Gibbs}\ \emph {et~al.}(2014)\citenamefont {Gibbs}, \citenamefont {Mark}, \citenamefont {Lee}, \citenamefont {Eslami}, \citenamefont {Schamel},\ and\ \citenamefont {Fischer}}]{Gibbs2014}%
  \BibitemOpen
  \bibfield  {author} {\bibinfo {author} {\bibfnamefont {J.~G.}\ \bibnamefont {Gibbs}}, \bibinfo {author} {\bibfnamefont {A.~G.}\ \bibnamefont {Mark}}, \bibinfo {author} {\bibfnamefont {T.-C.}\ \bibnamefont {Lee}}, \bibinfo {author} {\bibfnamefont {S.}~\bibnamefont {Eslami}}, \bibinfo {author} {\bibfnamefont {D.}~\bibnamefont {Schamel}}, \ and\ \bibinfo {author} {\bibfnamefont {P.}~\bibnamefont {Fischer}},\ }\bibfield  {title} {\enquote {\bibinfo {title} {Nanohelices by shadow growth},}\ }\href {\doibase 10.1039/C4NR00403E} {\bibfield  {journal} {\bibinfo  {journal} {Nanoscale}\ }\textbf {\bibinfo {volume} {6}},\ \bibinfo {pages} {9457--9466} (\bibinfo {year} {2014})}\BibitemShut {NoStop}%
\bibitem [{\citenamefont {Volkov}\ \emph {et~al.}(2019{\natexlab{b}})\citenamefont {Volkov}, \citenamefont {Rößler}, \citenamefont {Fassbender},\ and\ \citenamefont {Makarov}}]{Volkov_2019_JPhysD}%
  \BibitemOpen
  \bibfield  {author} {\bibinfo {author} {\bibfnamefont {O.~M.}\ \bibnamefont {Volkov}}, \bibinfo {author} {\bibfnamefont {U.~K.}\ \bibnamefont {Rößler}}, \bibinfo {author} {\bibfnamefont {J.}~\bibnamefont {Fassbender}}, \ and\ \bibinfo {author} {\bibfnamefont {D.}~\bibnamefont {Makarov}},\ }\bibfield  {title} {\enquote {\bibinfo {title} {Concept of artificial magnetoelectric materials via geometrically controlling curvilinear helimagnets},}\ }\href {\doibase 10.1088/1361-6463/ab2368} {\bibfield  {journal} {\bibinfo  {journal} {Journal of Physics D: Applied Physics}\ }\textbf {\bibinfo {volume} {52}},\ \bibinfo {pages} {345001} (\bibinfo {year} {2019}{\natexlab{b}})}\BibitemShut {NoStop}%
\bibitem [{\citenamefont {Trotta}\ \emph {et~al.}(2012)\citenamefont {Trotta}, \citenamefont {Atkinson}, \citenamefont {Plumhof}, \citenamefont {Zallo}, \citenamefont {Rezaev}, \citenamefont {Kumar}, \citenamefont {Baunack}, \citenamefont {Schröter}, \citenamefont {Rastelli},\ and\ \citenamefont {Schmidt}}]{Trotta2012}%
  \BibitemOpen
  \bibfield  {author} {\bibinfo {author} {\bibfnamefont {R.}~\bibnamefont {Trotta}}, \bibinfo {author} {\bibfnamefont {P.}~\bibnamefont {Atkinson}}, \bibinfo {author} {\bibfnamefont {J.~D.}\ \bibnamefont {Plumhof}}, \bibinfo {author} {\bibfnamefont {E.}~\bibnamefont {Zallo}}, \bibinfo {author} {\bibfnamefont {R.~O.}\ \bibnamefont {Rezaev}}, \bibinfo {author} {\bibfnamefont {S.}~\bibnamefont {Kumar}}, \bibinfo {author} {\bibfnamefont {S.}~\bibnamefont {Baunack}}, \bibinfo {author} {\bibfnamefont {J.~R.}\ \bibnamefont {Schröter}}, \bibinfo {author} {\bibfnamefont {A.}~\bibnamefont {Rastelli}}, \ and\ \bibinfo {author} {\bibfnamefont {O.~G.}\ \bibnamefont {Schmidt}},\ }\bibfield  {title} {\enquote {\bibinfo {title} {Nanomembrane quantum-light-emitting diodes integrated onto piezoelectric actuators},}\ }\href {\doibase https://doi.org/10.1002/adma.201200537} {\bibfield  {journal} {\bibinfo  {journal} {Advanced Materials}\ }\textbf {\bibinfo {volume} {24}},\ \bibinfo {pages} {2668--2672} (\bibinfo {year}
  {2012})}\BibitemShut {NoStop}%
\bibitem [{\citenamefont {Plumhof}\ \emph {et~al.}(2011)\citenamefont {Plumhof}, \citenamefont {K\ifmmode~\check{r}\else \v{r}\fi{}\'apek}, \citenamefont {Ding}, \citenamefont {J\"ons}, \citenamefont {Hafenbrak}, \citenamefont {Klenovsk\'y}, \citenamefont {Herklotz}, \citenamefont {D\"orr}, \citenamefont {Michler}, \citenamefont {Rastelli},\ and\ \citenamefont {Schmidt}}]{Plumhof2011}%
  \BibitemOpen
  \bibfield  {author} {\bibinfo {author} {\bibfnamefont {J.~D.}\ \bibnamefont {Plumhof}}, \bibinfo {author} {\bibfnamefont {V.}~\bibnamefont {K\ifmmode~\check{r}\else \v{r}\fi{}\'apek}}, \bibinfo {author} {\bibfnamefont {F.}~\bibnamefont {Ding}}, \bibinfo {author} {\bibfnamefont {K.~D.}\ \bibnamefont {J\"ons}}, \bibinfo {author} {\bibfnamefont {R.}~\bibnamefont {Hafenbrak}}, \bibinfo {author} {\bibfnamefont {P.}~\bibnamefont {Klenovsk\'y}}, \bibinfo {author} {\bibfnamefont {A.}~\bibnamefont {Herklotz}}, \bibinfo {author} {\bibfnamefont {K.}~\bibnamefont {D\"orr}}, \bibinfo {author} {\bibfnamefont {P.}~\bibnamefont {Michler}}, \bibinfo {author} {\bibfnamefont {A.}~\bibnamefont {Rastelli}}, \ and\ \bibinfo {author} {\bibfnamefont {O.~G.}\ \bibnamefont {Schmidt}},\ }\bibfield  {title} {\enquote {\bibinfo {title} {Strain-induced anticrossing of bright exciton levels in single self-assembled gaas/al${}_{x}$ga${}_{1\ensuremath{-}x}$as and in${}_{x}$ga${}_{1\ensuremath{-}x}$as/gaas quantum dots},}\ }\href {\doibase
  10.1103/PhysRevB.83.121302} {\bibfield  {journal} {\bibinfo  {journal} {Phys. Rev. B}\ }\textbf {\bibinfo {volume} {83}},\ \bibinfo {pages} {121302} (\bibinfo {year} {2011})}\BibitemShut {NoStop}%
\bibitem [{\citenamefont {Ding}\ \emph {et~al.}(2010)\citenamefont {Ding}, \citenamefont {Singh}, \citenamefont {Plumhof}, \citenamefont {Zander}, \citenamefont {K\ifmmode~\check{r}\else \v{r}\fi{}\'apek}, \citenamefont {Chen}, \citenamefont {Benyoucef}, \citenamefont {Zwiller}, \citenamefont {D\"orr}, \citenamefont {Bester}, \citenamefont {Rastelli},\ and\ \citenamefont {Schmidt}}]{Ding2010}%
  \BibitemOpen
  \bibfield  {author} {\bibinfo {author} {\bibfnamefont {F.}~\bibnamefont {Ding}}, \bibinfo {author} {\bibfnamefont {R.}~\bibnamefont {Singh}}, \bibinfo {author} {\bibfnamefont {J.~D.}\ \bibnamefont {Plumhof}}, \bibinfo {author} {\bibfnamefont {T.}~\bibnamefont {Zander}}, \bibinfo {author} {\bibfnamefont {V.}~\bibnamefont {K\ifmmode~\check{r}\else \v{r}\fi{}\'apek}}, \bibinfo {author} {\bibfnamefont {Y.~H.}\ \bibnamefont {Chen}}, \bibinfo {author} {\bibfnamefont {M.}~\bibnamefont {Benyoucef}}, \bibinfo {author} {\bibfnamefont {V.}~\bibnamefont {Zwiller}}, \bibinfo {author} {\bibfnamefont {K.}~\bibnamefont {D\"orr}}, \bibinfo {author} {\bibfnamefont {G.}~\bibnamefont {Bester}}, \bibinfo {author} {\bibfnamefont {A.}~\bibnamefont {Rastelli}}, \ and\ \bibinfo {author} {\bibfnamefont {O.~G.}\ \bibnamefont {Schmidt}},\ }\bibfield  {title} {\enquote {\bibinfo {title} {Tuning the exciton binding energies in single self-assembled $\mathrm{InGaAs}/\mathrm{GaAs}$ quantum dots by piezoelectric-induced biaxial stress},}\
  }\href {\doibase 10.1103/PhysRevLett.104.067405} {\bibfield  {journal} {\bibinfo  {journal} {Phys. Rev. Lett.}\ }\textbf {\bibinfo {volume} {104}},\ \bibinfo {pages} {067405} (\bibinfo {year} {2010})}\BibitemShut {NoStop}%
\bibitem [{\citenamefont {Zander}\ \emph {et~al.}(2009)\citenamefont {Zander}, \citenamefont {Herklotz}, \citenamefont {Kiravittaya}, \citenamefont {Benyoucef}, \citenamefont {Ding}, \citenamefont {Atkinson}, \citenamefont {Kumar}, \citenamefont {Plumhof}, \citenamefont {D\"{o}rr}, \citenamefont {Rastelli},\ and\ \citenamefont {Schmidt}}]{Zander2009}%
  \BibitemOpen
  \bibfield  {author} {\bibinfo {author} {\bibfnamefont {T.}~\bibnamefont {Zander}}, \bibinfo {author} {\bibfnamefont {A.}~\bibnamefont {Herklotz}}, \bibinfo {author} {\bibfnamefont {S.}~\bibnamefont {Kiravittaya}}, \bibinfo {author} {\bibfnamefont {M.}~\bibnamefont {Benyoucef}}, \bibinfo {author} {\bibfnamefont {F.}~\bibnamefont {Ding}}, \bibinfo {author} {\bibfnamefont {P.}~\bibnamefont {Atkinson}}, \bibinfo {author} {\bibfnamefont {S.}~\bibnamefont {Kumar}}, \bibinfo {author} {\bibfnamefont {J.~D.}\ \bibnamefont {Plumhof}}, \bibinfo {author} {\bibfnamefont {K.}~\bibnamefont {D\"{o}rr}}, \bibinfo {author} {\bibfnamefont {A.}~\bibnamefont {Rastelli}}, \ and\ \bibinfo {author} {\bibfnamefont {O.~G.}\ \bibnamefont {Schmidt}},\ }\bibfield  {title} {\enquote {\bibinfo {title} {Epitaxial quantum dots in stretchable optical microcavities},}\ }\href {\doibase 10.1364/OE.17.022452} {\bibfield  {journal} {\bibinfo  {journal} {Opt. Express}\ }\textbf {\bibinfo {volume} {17}},\ \bibinfo {pages} {22452--22461} (\bibinfo
  {year} {2009})}\BibitemShut {NoStop}%
\bibitem [{\citenamefont {Usadel}(1970)}]{Usadel1970}%
  \BibitemOpen
  \bibfield  {author} {\bibinfo {author} {\bibfnamefont {K.~D.}\ \bibnamefont {Usadel}},\ }\bibfield  {title} {\enquote {\bibinfo {title} {Generalized diffusion equation for superconducting alloys},}\ }\href {\doibase 10.1103/PhysRevLett.25.507} {\bibfield  {journal} {\bibinfo  {journal} {Phys. Rev. Lett.}\ }\textbf {\bibinfo {volume} {25}},\ \bibinfo {pages} {507--509} (\bibinfo {year} {1970})}\BibitemShut {NoStop}%
\bibitem [{\citenamefont {Kuprianov}\ and\ \citenamefont {Lukichev}()}]{KuprianovLukichev1988}%
  \BibitemOpen
  \bibfield  {author} {\bibinfo {author} {\bibfnamefont {M.~Y.}\ \bibnamefont {Kuprianov}}\ and\ \bibinfo {author} {\bibfnamefont {V.}~\bibnamefont {Lukichev}},\ }\bibfield  {title} {\enquote {\bibinfo {title} {Influence of boundary transparency on the critical current of dirty ss’s structures},}\ }\href {http://jetp.ras.ru/cgi-bin/dn/e_067_06_1163.pdf} {\bibinfo  {journal} {Zh. Eksp. Teor. Fiz (94) 139 (1988); Sov. Phys. JETP 67 (6) 1163-1168 (1988)}\ }\BibitemShut {NoStop}%
\bibitem [{\citenamefont {Ortix}(2015)}]{ortix2015quantum}%
  \BibitemOpen
\bibfield  {journal} {  }\bibfield  {author} {\bibinfo {author} {\bibfnamefont {C.}~\bibnamefont {Ortix}},\ }\bibfield  {title} {\enquote {\bibinfo {title} {Quantum mechanics of a spin-orbit coupled electron constrained to a space curve},}\ }\href {\doibase 10.1103/PhysRevB.91.245412} {\bibfield  {journal} {\bibinfo  {journal} {Phys. Rev. B}\ }\textbf {\bibinfo {volume} {91}},\ \bibinfo {pages} {245412} (\bibinfo {year} {2015})}\BibitemShut {NoStop}%
\bibitem [{\citenamefont {Schopohl}\ and\ \citenamefont {Maki}(1995)}]{Schopohl1995}%
  \BibitemOpen
  \bibfield  {author} {\bibinfo {author} {\bibfnamefont {N.}~\bibnamefont {Schopohl}}\ and\ \bibinfo {author} {\bibfnamefont {K.}~\bibnamefont {Maki}},\ }\bibfield  {title} {\enquote {\bibinfo {title} {Quasiparticle spectrum around a vortex line in a d-wave superconductor},}\ }\href {\doibase 10.1103/PhysRevB.52.490} {\bibfield  {journal} {\bibinfo  {journal} {Phys. Rev. B}\ }\textbf {\bibinfo {volume} {52}},\ \bibinfo {pages} {490--493} (\bibinfo {year} {1995})}\BibitemShut {NoStop}%
\bibitem [{\citenamefont {Ouassou}(2018)}]{Geneus}%
  \BibitemOpen
  \bibfield  {author} {\bibinfo {author} {\bibfnamefont {J.}~\bibnamefont {Ouassou}},\ }\bibfield  {title} {\enquote {\bibinfo {title} {Geneus},}\ }\href {https:// github.com/jabirali/geneus} {\bibfield  {journal} {\bibinfo  {journal} {GitHub repository}\ } (\bibinfo {year} {2018})}\BibitemShut {NoStop}%
\bibitem [{\citenamefont {Zgirski}\ and\ \citenamefont {Arutyunov}(2007)}]{Zgirski2007}%
  \BibitemOpen
  \bibfield  {author} {\bibinfo {author} {\bibfnamefont {M.}~\bibnamefont {Zgirski}}\ and\ \bibinfo {author} {\bibfnamefont {K.~Y.}\ \bibnamefont {Arutyunov}},\ }\bibfield  {title} {\enquote {\bibinfo {title} {Experimental limits of the observation of thermally activated phase-slip mechanism in superconducting nanowires},}\ }\href {\doibase 10.1103/PhysRevB.75.172509} {\bibfield  {journal} {\bibinfo  {journal} {Phys. Rev. B}\ }\textbf {\bibinfo {volume} {75}},\ \bibinfo {pages} {172509} (\bibinfo {year} {2007})}\BibitemShut {NoStop}%
\bibitem [{\citenamefont {Arutyunov}, \citenamefont {Golubev},\ and\ \citenamefont {Zaikin}(2008)}]{Arutyunov2008}%
  \BibitemOpen
  \bibfield  {author} {\bibinfo {author} {\bibfnamefont {K.}~\bibnamefont {Arutyunov}}, \bibinfo {author} {\bibfnamefont {D.}~\bibnamefont {Golubev}}, \ and\ \bibinfo {author} {\bibfnamefont {A.}~\bibnamefont {Zaikin}},\ }\bibfield  {title} {\enquote {\bibinfo {title} {Superconductivity in one dimension},}\ }\href {\doibase https://doi.org/10.1016/j.physrep.2008.04.009} {\bibfield  {journal} {\bibinfo  {journal} {Physics Reports}\ }\textbf {\bibinfo {volume} {464}},\ \bibinfo {pages} {1--70} (\bibinfo {year} {2008})}\BibitemShut {NoStop}%
\bibitem [{\citenamefont {Tkachenko}\ \emph {et~al.}(2012)\citenamefont {Tkachenko}, \citenamefont {Kuchko}, \citenamefont {Dvornik},\ and\ \citenamefont {Kruglyak}}]{Tkachenko2012}%
  \BibitemOpen
  \bibfield  {author} {\bibinfo {author} {\bibfnamefont {V.~S.}\ \bibnamefont {Tkachenko}}, \bibinfo {author} {\bibfnamefont {A.~N.}\ \bibnamefont {Kuchko}}, \bibinfo {author} {\bibfnamefont {M.}~\bibnamefont {Dvornik}}, \ and\ \bibinfo {author} {\bibfnamefont {V.~V.}\ \bibnamefont {Kruglyak}},\ }\bibfield  {title} {\enquote {\bibinfo {title} {{Propagation and scattering of spin waves in curved magnonic waveguides}},}\ }\href {\doibase 10.1063/1.4757994} {\bibfield  {journal} {\bibinfo  {journal} {Applied Physics Letters}\ }\textbf {\bibinfo {volume} {101}},\ \bibinfo {pages} {152402} (\bibinfo {year} {2012})}\BibitemShut {NoStop}%
\bibitem [{\citenamefont {Tkachenko}, \citenamefont {Kuchko},\ and\ \citenamefont {Kruglyak}(2013)}]{Tkachenko2013}%
  \BibitemOpen
  \bibfield  {author} {\bibinfo {author} {\bibfnamefont {V.~S.}\ \bibnamefont {Tkachenko}}, \bibinfo {author} {\bibfnamefont {A.~N.}\ \bibnamefont {Kuchko}}, \ and\ \bibinfo {author} {\bibfnamefont {V.~V.}\ \bibnamefont {Kruglyak}},\ }\bibfield  {title} {\enquote {\bibinfo {title} {{An effect of the curvature induced anisotropy on the spectrum of spin waves in a curved magnetic nanowire}},}\ }\href {\doibase 10.1063/1.4792133} {\bibfield  {journal} {\bibinfo  {journal} {Low Temperature Physics}\ }\textbf {\bibinfo {volume} {39}},\ \bibinfo {pages} {163--166} (\bibinfo {year} {2013})}\BibitemShut {NoStop}%
\bibitem [{\citenamefont {Slastikov}\ and\ \citenamefont {Sonnenberg}(2011)}]{Slastikov2011}%
  \BibitemOpen
  \bibfield  {author} {\bibinfo {author} {\bibfnamefont {V.~V.}\ \bibnamefont {Slastikov}}\ and\ \bibinfo {author} {\bibfnamefont {C.}~\bibnamefont {Sonnenberg}},\ }\bibfield  {title} {\enquote {\bibinfo {title} {{Reduced models for ferromagnetic nanowires}},}\ }\href {\doibase 10.1093/imamat/hxr019} {\bibfield  {journal} {\bibinfo  {journal} {IMA Journal of Applied Mathematics}\ }\textbf {\bibinfo {volume} {77}},\ \bibinfo {pages} {220--235} (\bibinfo {year} {2011})}\BibitemShut {NoStop}%
\bibitem [{\citenamefont {Kundys}(2015)}]{Kundys2015}%
  \BibitemOpen
  \bibfield  {author} {\bibinfo {author} {\bibfnamefont {B.}~\bibnamefont {Kundys}},\ }\bibfield  {title} {\enquote {\bibinfo {title} {Photostrictive materials},}\ }\href {\doibase 10.1063/1.4905505} {\bibfield  {journal} {\bibinfo  {journal} {Applied Physics Reviews}\ }\textbf {\bibinfo {volume} {2}},\ \bibinfo {pages} {011301} (\bibinfo {year} {2015})}\BibitemShut {NoStop}%
\bibitem [{\citenamefont {Matzen}\ \emph {et~al.}(2019)\citenamefont {Matzen}, \citenamefont {Guillemot}, \citenamefont {Maroutian}, \citenamefont {Patel}, \citenamefont {Wen}, \citenamefont {DiChiara}, \citenamefont {Agnus}, \citenamefont {Shpyrko}, \citenamefont {Fullerton}, \citenamefont {Ravelosona}, \citenamefont {Lecoeur},\ and\ \citenamefont {Kukreja}}]{Matzen2019}%
  \BibitemOpen
  \bibfield  {author} {\bibinfo {author} {\bibfnamefont {S.}~\bibnamefont {Matzen}}, \bibinfo {author} {\bibfnamefont {L.}~\bibnamefont {Guillemot}}, \bibinfo {author} {\bibfnamefont {T.}~\bibnamefont {Maroutian}}, \bibinfo {author} {\bibfnamefont {S.~K.~K.}\ \bibnamefont {Patel}}, \bibinfo {author} {\bibfnamefont {H.}~\bibnamefont {Wen}}, \bibinfo {author} {\bibfnamefont {A.~D.}\ \bibnamefont {DiChiara}}, \bibinfo {author} {\bibfnamefont {G.}~\bibnamefont {Agnus}}, \bibinfo {author} {\bibfnamefont {O.~G.}\ \bibnamefont {Shpyrko}}, \bibinfo {author} {\bibfnamefont {E.~E.}\ \bibnamefont {Fullerton}}, \bibinfo {author} {\bibfnamefont {D.}~\bibnamefont {Ravelosona}}, \bibinfo {author} {\bibfnamefont {P.}~\bibnamefont {Lecoeur}}, \ and\ \bibinfo {author} {\bibfnamefont {R.}~\bibnamefont {Kukreja}},\ }\bibfield  {title} {\enquote {\bibinfo {title} {Tuning ultrafast photoinduced strain in ferroelectric-based devices},}\ }\href {\doibase https://doi.org/10.1002/aelm.201800709} {\bibfield  {journal} {\bibinfo
  {journal} {Advanced Electronic Materials}\ }\textbf {\bibinfo {volume} {5}},\ \bibinfo {pages} {1800709} (\bibinfo {year} {2019})}\BibitemShut {NoStop}%
\bibitem [{\citenamefont {Guillemeney}\ \emph {et~al.}(2022)\citenamefont {Guillemeney}, \citenamefont {Lermusiaux}, \citenamefont {Landaburu}, \citenamefont {Wagnon},\ and\ \citenamefont {Ab{\'e}cassis}}]{Guillemeney2022}%
  \BibitemOpen
  \bibfield  {author} {\bibinfo {author} {\bibfnamefont {L.}~\bibnamefont {Guillemeney}}, \bibinfo {author} {\bibfnamefont {L.}~\bibnamefont {Lermusiaux}}, \bibinfo {author} {\bibfnamefont {G.}~\bibnamefont {Landaburu}}, \bibinfo {author} {\bibfnamefont {B.}~\bibnamefont {Wagnon}}, \ and\ \bibinfo {author} {\bibfnamefont {B.}~\bibnamefont {Ab{\'e}cassis}},\ }\bibfield  {title} {\enquote {\bibinfo {title} {Curvature and self-assembly of semi-conducting nanoplatelets},}\ }\href {\doibase 10.1038/s42004-021-00621-z} {\bibfield  {journal} {\bibinfo  {journal} {Communications Chemistry}\ }\textbf {\bibinfo {volume} {5}},\ \bibinfo {pages} {7} (\bibinfo {year} {2022})}\BibitemShut {NoStop}%
\bibitem [{\citenamefont {Park}\ and\ \citenamefont {Shrout}(1997)}]{Park1997}%
  \BibitemOpen
  \bibfield  {author} {\bibinfo {author} {\bibfnamefont {S.-E.}\ \bibnamefont {Park}}\ and\ \bibinfo {author} {\bibfnamefont {T.~R.}\ \bibnamefont {Shrout}},\ }\bibfield  {title} {\enquote {\bibinfo {title} {{Ultrahigh strain and piezoelectric behavior in relaxor based ferroelectric single crystals}},}\ }\href {\doibase 10.1063/1.365983} {\bibfield  {journal} {\bibinfo  {journal} {Journal of Applied Physics}\ }\textbf {\bibinfo {volume} {82}},\ \bibinfo {pages} {1804--1811} (\bibinfo {year} {1997})}\BibitemShut {NoStop}%
\bibitem [{\citenamefont {Sheka}, \citenamefont {Kravchuk},\ and\ \citenamefont {Gaididei}(2015)}]{Sheka2015}%
  \BibitemOpen
  \bibfield  {author} {\bibinfo {author} {\bibfnamefont {D.~D.}\ \bibnamefont {Sheka}}, \bibinfo {author} {\bibfnamefont {V.~P.}\ \bibnamefont {Kravchuk}}, \ and\ \bibinfo {author} {\bibfnamefont {Y.}~\bibnamefont {Gaididei}},\ }\bibfield  {title} {\enquote {\bibinfo {title} {Curvature effects in statics and dynamics of low dimensional magnets},}\ }\href {\doibase 10.1088/1751-8113/48/12/125202} {\bibfield  {journal} {\bibinfo  {journal} {Journal of Physics A: Mathematical and Theoretical}\ }\textbf {\bibinfo {volume} {48}},\ \bibinfo {pages} {125202} (\bibinfo {year} {2015})}\BibitemShut {NoStop}%
\bibitem [{\citenamefont {Sheka}\ \emph {et~al.}(2015)\citenamefont {Sheka}, \citenamefont {Kravchuk}, \citenamefont {Yershov},\ and\ \citenamefont {Gaididei}}]{Sheka2015a}%
  \BibitemOpen
  \bibfield  {author} {\bibinfo {author} {\bibfnamefont {D.~D.}\ \bibnamefont {Sheka}}, \bibinfo {author} {\bibfnamefont {V.~P.}\ \bibnamefont {Kravchuk}}, \bibinfo {author} {\bibfnamefont {K.~V.}\ \bibnamefont {Yershov}}, \ and\ \bibinfo {author} {\bibfnamefont {Y.}~\bibnamefont {Gaididei}},\ }\bibfield  {title} {\enquote {\bibinfo {title} {Torsion-induced effects in magnetic nanowires},}\ }\href {\doibase 10.1103/PhysRevB.92.054417} {\bibfield  {journal} {\bibinfo  {journal} {Physical Review B}\ }\textbf {\bibinfo {volume} {92}},\ \bibinfo {pages} {054417} (\bibinfo {year} {2015})}\BibitemShut {NoStop}%
\bibitem [{\citenamefont {Ozeri}\ \emph {et~al.}(2023)\citenamefont {Ozeri}, \citenamefont {Devidas}, \citenamefont {Alpern}, \citenamefont {Persky}, \citenamefont {Bjorlig}, \citenamefont {Sukenik}, \citenamefont {Yochelis}, \citenamefont {Di~Bernardo}, \citenamefont {Kalisky}, \citenamefont {Millo},\ and\ \citenamefont {Paltiel}}]{Ozeri2023}%
  \BibitemOpen
  \bibfield  {author} {\bibinfo {author} {\bibfnamefont {M.}~\bibnamefont {Ozeri}}, \bibinfo {author} {\bibfnamefont {T.}~\bibnamefont {Devidas}}, \bibinfo {author} {\bibfnamefont {H.}~\bibnamefont {Alpern}}, \bibinfo {author} {\bibfnamefont {E.}~\bibnamefont {Persky}}, \bibinfo {author} {\bibfnamefont {A.~V.}\ \bibnamefont {Bjorlig}}, \bibinfo {author} {\bibfnamefont {N.}~\bibnamefont {Sukenik}}, \bibinfo {author} {\bibfnamefont {S.}~\bibnamefont {Yochelis}}, \bibinfo {author} {\bibfnamefont {A.}~\bibnamefont {Di~Bernardo}}, \bibinfo {author} {\bibfnamefont {B.}~\bibnamefont {Kalisky}}, \bibinfo {author} {\bibfnamefont {O.}~\bibnamefont {Millo}}, \ and\ \bibinfo {author} {\bibfnamefont {Y.}~\bibnamefont {Paltiel}},\ }\bibfield  {title} {\enquote {\bibinfo {title} {Scanning {SQUID} imaging of reduced superconductivity due to the effect of chiral molecule islands adsorbed on {N}b},}\ }\href {\doibase https://doi.org/10.1002/admi.202201899} {\bibfield  {journal} {\bibinfo  {journal} {Advanced Materials
  Interfaces}\ }\textbf {\bibinfo {volume} {10}},\ \bibinfo {pages} {2201899} (\bibinfo {year} {2023})}\BibitemShut {NoStop}%
\end{thebibliography}%


\end{document}